\documentclass[aip,reprint,amsmath,amssymb]{revtex4-2}
\usepackage{color}
\usepackage{graphicx}
\usepackage{amsmath}
\usepackage{subcaption}
\usepackage{amssymb}
\usepackage{lineno}
\usepackage{multirow}
\usepackage{natbib}

\graphicspath{{figs/}}

\usepackage{placeins}

\begin{document}
	\title{Plasma transport by carrier electrons: Nernst advection beyond the heat-flow analogy due to kinetic stress}

	\author{C. A. Walsh}
	\email{walsh34@llnl.gov}
	\affiliation{Lawrence Livermore National Laboratory, 7000 East Avenue, Livermore, CA 94550}

	\date{\today}

	%\linenumbers

	\begin{abstract}
	Nernst advection of magnetic fields down temperature gradients is lowered by kinetic stress from the $\underline{\underline{f}}_2$ electron distribution harmonic, which allows fluxes to stream past one another without isotropizing their lateral momentum. The long-held tie between heat-flow and Nernst advection is disrupted. When the distance between electron collisions is comparable to the magnetic field length-scale $L_B$, the kinetic stress affects magnetic transport in plasmas otherwise regarded as classical. This is from carrier electrons being deflected back and forth over the course of their trajectory. The field structure Knudsen number is defined: $Kn_B=14.8\frac{\lambda_{ei}}{L_B}\frac{Z+0.40}{Z+4.86}$, with Nernst velocity halved by kinetic effects at $Kn_B=1$. The kinetic stress is relocalized by magnetization, following the relocalization of heat-flux. Righi-Leduc heat-flow is found to be similarly suppressed by kinetic stress. The linear results of this paper are found to faithfully represent the non-linear evolution of magnetic field at a heat-front, indicating that low-$Z$ plasmas throughout magnetized-HED are demagnetized more slowly than previously anticipated.
	\end{abstract}
	\maketitle

	Electron thermal conduction is a dominant energy transport mechanism in laser-driven plasma experiments \cite{spitzerTransportPhenomenaCompletely1953}. An analogous process occurs in the induction equation: Nernst advection of magnetic fields down temperature gradients\cite{hainesHeatFluxEffects1986}. Across magnetized high-energy-density physics this process has been found to demagnetize hot plasmas \cite{slutzHighGainMagnetizedInertial2012a,farmerSimulationSelfgeneratedMagnetic2017,walshSelfGeneratedMagneticFields2017,hillEnhancementPressurePerturbations2017,leal2025,bissellNernstAdvectionFieldgenerating2015,chittendenNernstEttinghausenEffects1993,garcia-rubioMagneticFluxConservation2018,hataDynamicsSelfGeneratedMagnetic2006,igumenshchevSelfgeneratedMagneticFields2014,nishiguchiNernstEffectLaserproduced1985,velikovichMagneticFluxHeat2015a,2021}, with recent experiments confirming existence of the effect\cite{arran2025}.

	The tie between heat-flux and Nernst in the classical regime has been noted for over 40 years\cite{hainesHeatFluxEffects1986}. More recently, theoretical and computational work has suggested the link between the two processes continues into the kinetic regime, with heat-flux and Nernst suppressed by the same fraction due to delocalization of the electron distribution tail \cite{joglekarKineticModelingNernst2016,brodrickTestingNonlocalModels2017,brodrickIncorporatingKineticEffects2018,sherlockSuppressionBiermannBattery2020,walshKineticCorrectionsHeatflow2024}. This paper shows that those studies overlooked a transport term, whose local form is Liu \textit{et al.}'s 'heat-flux viscosity'\cite{liu2015}, previously identified as a dissipation mechanism in reconnection. This process smears the Nernst flux in the kinetic regime, breaking the link between heat-flux and Nernst advection, and resulting in more magnetized plasmas than would otherwise be expected.

	In the classical regime, the Nernst term of the induction equation can be written as an advection of the magnetic field $\mathbf{B}$ with velocity $\mathbf{v}_{N}$:\cite{braginskiiTransportProcessesPlasma1965,walshExtendedmagnetohydrodynamicsDensePlasmas2020}
	\begin{equation}
		\mathbf{v}_{N} = - \frac{\tau_B}{m_e}\left(\gamma_{\bot}^c \nabla T_e + \gamma_{\wedge}^c\, \mathbf{\hat{b}} \times \nabla T_e\right) \label{eq:nernst}
	\end{equation}
	where the second term is cross-gradient Nernst, which advects field perpendicular to both the temperature gradient and the magnetic field direction $\mathbf{\hat{b}}$. Throughout this paper the magnetic field is taken perpendicular to the temperature gradient.  $\gamma_{\bot}^c$ and $\gamma_{\wedge}^c$ are dimensionless coefficients that depend only on the ionization $Z$ and Hall parameter $\chi$. $\tau_B$ is the collision time defined by Braginskii\cite{braginskiiTransportProcessesPlasma1965}.

	The heat flux is naturally diffusive, but it can be written as a heat-flow velocity for comparison with equation \ref{eq:nernst}:
	\begin{equation}
		\mathbf{v}_q = -\frac{2}{5}\,\frac{\tau_B}{m_e}\left(\kappa_\perp^c \nabla T_e + \kappa_{\wedge}^c\, \mathbf{\hat{b}} \times \nabla T_e\right) \label{eq:heat}
	\end{equation}

\begin{figure}[tb]
		\includegraphics[width=\linewidth]{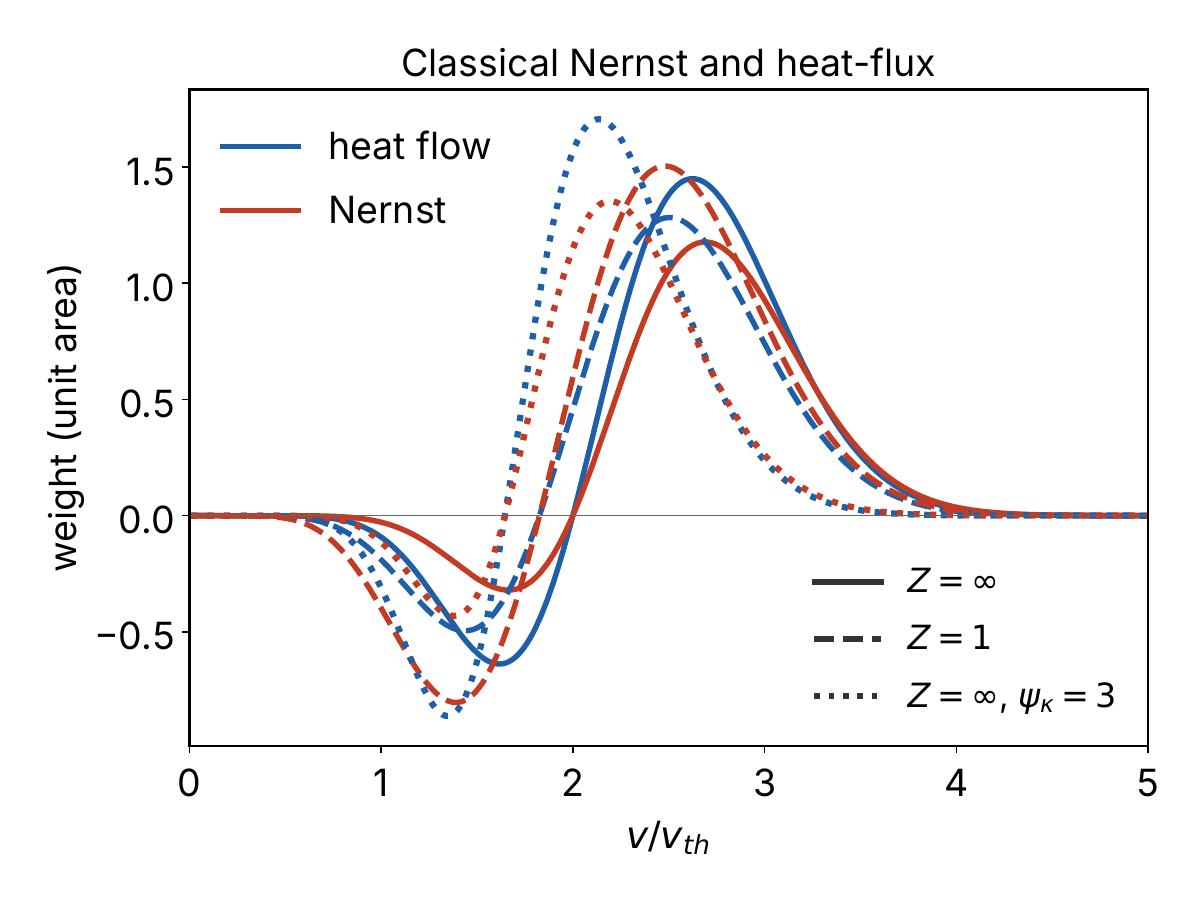}
		\caption{\label{fig:classical_weights} The electrons that carry Nernst and heat-flux in a classical plasma. e-e collisions (low $Z$) and magnetization ($\psi_\kappa > 0$) affect the carrier electrons similarly, which is why the literature orthodoxy is that heat-flux and Nernst should be equally susceptible to kinetic suppression\cite{walshKineticCorrectionsHeatflow2024}. However, the kinetic stress severs this relationship. }
	\end{figure}

	Nernst ($\gamma_{\bot}^c$) and perpendicular heat ($\kappa_{\bot}^c$) are also carried by similar electron populations. Figure \ref{fig:classical_weights} shows how each process's flux is dependent on each electron velocity $v$. The solid lines are heat-flux and Nernst in the Lorentz limit ($Z=\infty$), where electron-electron (e-e) collisions can be ignored; carrier velocity peak for Nernst is $2.69v_{th}$ compared with $2.62v_{th}$ for heat-flux. $v_{th}=\sqrt{2T_e/m_e}$ is the velocity of thermal electrons.  The dashed lines in figure \ref{fig:classical_weights} are for a classical $Z=1$ plasma, where e-e collisions lower the carrier velocities to $2.48v_{th}$ and $2.51v_{th}$. The dotted lines are for a plasma magnetized to a carrier Hall parameter\cite{walsh_paper0} of $\psi_\kappa = 3$ (corresponding to $\chi=0.3$ for $Z=\infty$), giving $2.20v_{th}$ for Nernst and $2.13v_{th}$ for heat-flow. The carrier velocities of heat-flux and Nernst move together, suggesting that they are equally susceptible to kinetic effects\cite{walsh_paper0}.

	Recent work\cite{walsh_paper1} used $>10^5$ 1D Vlasov-Fokker Planck (VFP) simulations to map the kinetic suppression of heat-flux in the linear regime across temperature perturbation wavenumber ($k_T$), ionization ($Z$) and magnetization ($\chi$). A key outcome was measurement of the delocalization length ($\lambda_d=\lambda_{ei}Z/\sqrt{Z+4.45}$), which results in all plasma ionizations following the same kinetic heat-flux suppression function $\Gamma_q(k_T\lambda_{d})$. The heat-flux was found to be relocalized by magnetization through the delocalization Hall parameter ($\psi_d = 13.4 \chi Z/(Z+4.45)$), giving a magnetized delocalization length $\lambda_{d\bot} = \lambda_d/(1+\psi_d^{1.6})$.\cite{walsh_paper1}

\begin{figure}[tb]
		\includegraphics[width=\linewidth]{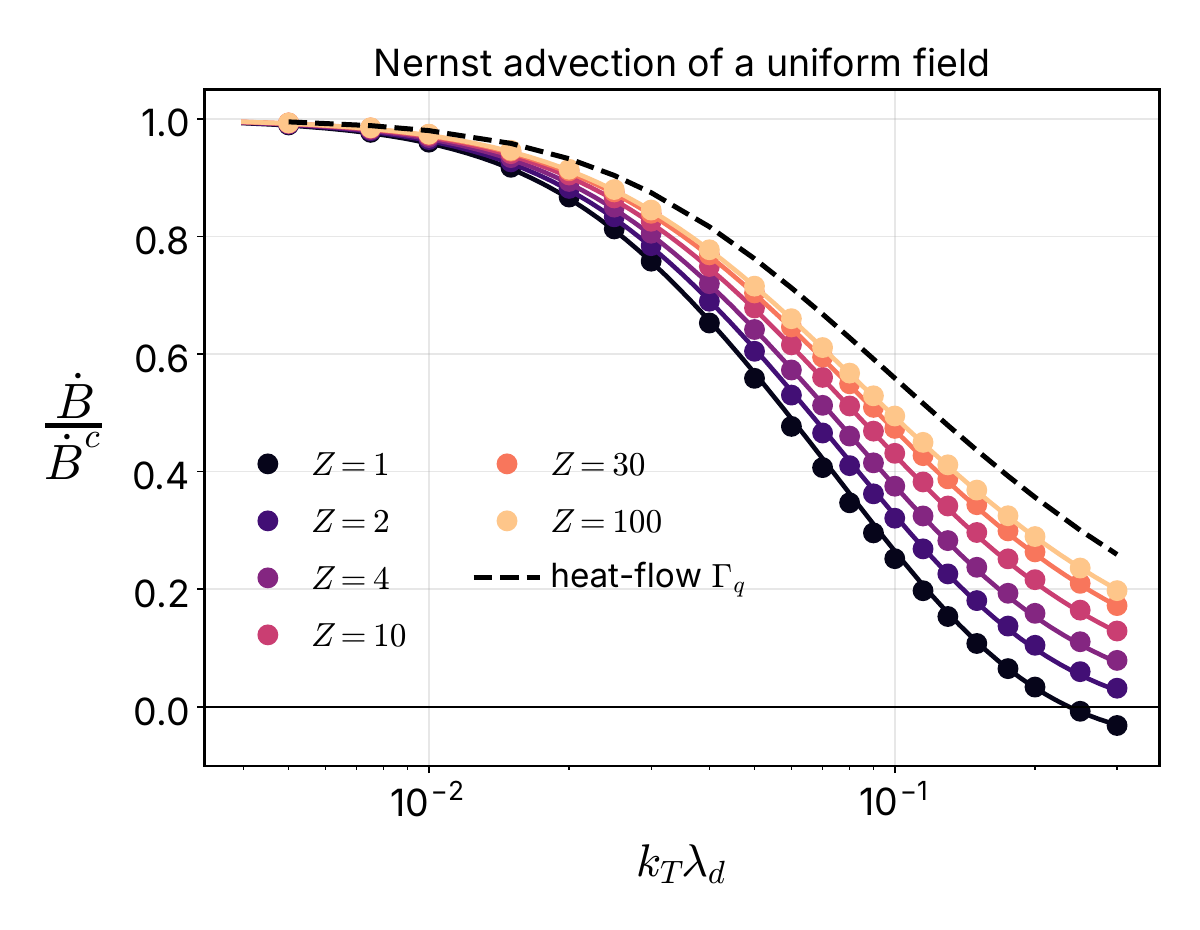}
		\caption{\label{fig:nernst_sup} Kinetic suppression of magnetic field advection beyond heat-flux suppression ($\Gamma_q$), particularly for low-$Z$ plasmas. Heat-flow and field advection are measured simultaneously from 1D VFP simulations with small uniform magnetic field\cite{walsh_paper1}. This breaks the long-established tie between heat-flux and Nernst\cite{hainesHeatFluxEffects1986}.% resolved 19 Nov: figure x-axis now k_T lambda_d; superscript k removed from Bdot
	}
	\end{figure}

	Figure \ref{fig:nernst_sup} is the central motivation of this paper, showing a break from the long-held wisdom that Nernst and heat-flux are intrinsically linked. It uses the same simulations presented in the heat-flux delocalization work\cite{walsh_paper1}; Nernst and heat-flux are measured simultaneously. The kinetic rate of change of magnetic field strength normalized to classical ($\dot{B}/\dot{B}^c$) is plotted against the non-locality metric $k_T \lambda_d$, in the weak magnetic field limit. For these 1D simulations with uniform perpendicular field and stationary ions, the classical $\dot{B}^c$ is entirely due to Nernst moving magnetic fields down the temperature gradient.

	The dashed line in figure \ref{fig:nernst_sup} shows the heat-flux suppression $\Gamma_q$, which is independent of ionization. $\dot{B}$ at $Z=100$ is as expected from Nernst: marginally more suppressed than heat-flux, which is attributed to it being carried by a marginally higher velocity electron population. However, Nernst becomes more suppressed than heat-flux for lower $Z$, with the non-locality parameter $k_T \lambda_d$ unable to collapse the response of magnetic field to Nernst.  In fact, for $Z=1$, the plot suggests Nernst \emph{reverses sign}, pushing magnetic field \emph{up} the temperature gradient. What is going on?

	The answer comes from the kinetic stress, which will be shown across the following narrative arc. Section \ref{sec:ohm_f1} considers Ohm's law when the spherical harmonic expansion is truncated at $\mathbf{f}_1$, which removes the kinetic stress. Section \ref{sec:ohm_f2} then demonstrates the origin of anisotropic temperatures and stress at sharp temperature gradients, resulting in an extended Ohm's law in section \ref{sec:ohm_full}. Section \ref{sec:vfp} outlines the new  1D VFP simulations used to characterize magnetic field advection, which include both a thermal ($k_T$) and magnetic ($k_B$) perturbation.

	Section \ref{sec:weak_field} then studies these new simulations in the limit of weak magnetic field, separating out 'pure Nernst' from stress effects. While pure Nernst does indeed follow the kinetic suppression of heat-flux, the importance of including stress effects is emphasized, especially when the characteristic length-scale of the magnetic field structure is less than the delocalization length. Section \ref{sec:mag} extends this understanding to magnetized plasmas, with the kinetic stress being relocalized at a similar rate to heat-flow non-locality. Section \ref{sec:nernst_heat} revisits the heat-flux analogy for Nernst, arguing that Nernst shares its origin with Righi-Leduc heat-flow, with both resulting from electron flux being rotated by the magnetic field. This leads to finding that Righi-Leduc is susceptible to the same kinetic stress suppression as Nernst. Finally, section \ref{sec:implication} tests the linear results on a non-linear heat-front; demagnetization is reduced as linear theory predicts.

	Table \ref{tab:nernst_terms} summarizes the transport processes in this paper for a uniform field.

	Appendix \ref{sec:aniso_fits} provides fits to the bulk temperature anisotropy and shear data. Appendix \ref{sec:app_vfp_time} compares the quasi-static linear mode VFP data of this paper against full time-dependent VFP calculations, validating the approach. Appendix \ref{sec:app_closure} extends the fits of table \ref{tab:nernst_terms} to non-uniform magnetic fields.

% Nernst paper: table of the terms in the advecting field, draft 2 (17 Nov 2026).
% Decaying definition throughout (as Paper 1). Small-k coefficients: nernst_table/smallk.py on the weak-field two-mode set
% (L = 6), lambda_d^2 units. Names: S_F = flux-driven shear (old S_A, coefficient F), S_A = anisotropy-driven shear
% (old S_R, coefficient A). Hall parameters of the stress rows: single-Hall fits from the v6 closure work (mag5);
% S_f0 constants from MAG_CLOSURE_HYBRID_15NOV.md; Nernst total from D3. Carrier speeds: Lorentz (Z -> infinity).
\begin{table*}[tb]
	\caption{The processes affecting magnetic field transport,
		with the heat-flow for comparison. The
		small-wavenumber ($k$) coefficients: $1-\Gamma_X \simeq c\,(k_T\lambda_d)^2$, $S_F \simeq c\,(k\lambda_d)^2$ and
		$S_A \simeq c\,(k_T\lambda_d)(k\lambda_d)$, where $k_T$ is the wavenumber of the temperature structure and $k$ that of
		the combined structure ($k = k_T + k_B$). $\Gamma_X$ is the suppression function for process $X$, with $\Gamma_X = 1$ being unsuppressed. $P(K) = (1 + 27.29K)/(1 + 27.29K + 294.5K^2)$ is the suppression function measured for heat-flux\cite{walsh_paper1}. Carrier speed: where the
		term's per-speed weight peaks}
	\label{tab:nernst_terms}
	\begin{ruledtabular}
	\renewcommand{\arraystretch}{1.5}\small\setlength{\tabcolsep}{3pt}
	\begin{tabular}{lllcccc}
		Term & Driven by & Depends & small-$k$ $c$ & finite $k$ & Hall parameter & carrier \\
		& & on & $Z=1$ / $100$ & & & speed \\
		\colrule
		Heat flow, $\Gamma_q$     & $\nabla T_e$             & $k_T$        & 224 / 248 & $P(k_T\lambda_d)$       & $\psi_d$                                & 2.62 \\
		Nernst, $\Gamma_N$        & $\nabla T_e$             & $k_T$        & 231 / 301 & $P(1.10\,k_T\lambda_d)$ & $0.78\,\psi_d$                          & 2.69 \\
		Flux shear, $S_F$         & rotated heat flow        & $k_T$, $k_B$ &  90 / 4.9 & filtered$^\dagger$      & $\frac{22.2\,Z\chi}{Z+11.5}$, $p = 2.0$ & 3.3 \\
		Anisotropy shear, $S_A$   & rotated normal stress    & $k_T$, $k_B$ & 138 / 3.9 & filtered$^\dagger$      & $\frac{15.4\,Z\chi}{Z+6.06}$, $p = 2.1$ & 3.3 \\
		$f_0$ shape, $S_{f0}$     & $B$-induced $\delta f_0$ & $k_T$, $k_B$ & $0$ at $B\to0$ & filtered$^\dagger$ & ---                                  & --- \\
		\colrule
		Nernst total, uniform $B$ & all, at $k_B=0$          & $k_T$        & 464 / 310 & $1 - S_N[1 - P(k_T\lambda_N)]$ & hybrid$^\dagger$                 & --- \\
	\end{tabular}
\end{ruledtabular}
	\raggedright\footnotesize
	$^\dagger$ Closures in appendix~\ref{sec:app_closure}. Nernst total: $\lambda_N = 1.15\,\frac{Z+0.67}{Z+0.30}\,\lambda_d$, $S_N = 1.03\,\frac{Z+6.32}{Z+4.81}$; its small-$k$ coefficient
	is close to the sum of the rows above it.
\end{table*}

\section{Kinetic induction equation}
\label{sec:ohm}

Spherical harmonics are used to expand the electron distribution\cite{bell1981,shkarofskyip1966}: $f_0(v)$, $\mathbf{f}_1(v)$, $\underline{\underline{f}}_2(v)$, etc, where $v$ is the electron speed. $f_0$ contains the bulk plasma properties such as temperature and density. $\mathbf{f}_1(v)$ contains all fluxes; $f_{1x}(v)>0$ says that more electrons at speed $v$ are travelling in positive $x$ than negative $x$.

Considering this is an article on plasma transport, it would seem reasonable to truncate the harmonic expansion at $\mathbf{f}_1$, as $\underline{\underline{f}}_2(v)$ carries no flux. Section \ref{sec:ohm_f1} takes this route, with Nernst the only magnetic transport process for the assumed conditions: 1D with immobile ions and no current. However, the higher harmonics still feed back onto $\mathbf{f}_1$, thereby affecting the flux. Section \ref{sec:ohm_f2} introduces $\underline{\underline{f}}_2$, including the mechanisms by which temperature anisotropy ($T_{xx} \ne T_{yy}$) and stress ($T_{xy}>0$) are driven. Section \ref{sec:ohm_full} then forms a kinetic Ohm's law including the heat-flux viscosity, showing how it resists Nernst advection.

%---------------------------------------------------------------------------------------------------------------
\subsection{Kinetic origin of Nernst at $\mathbf f_1$ truncation}
\label{sec:ohm_f1}

	A steady-state ($\partial \mathbf{f}_1/\partial t = 0$) plasma with ions at rest ($v_{plasma}=0$) and no current ($\mathbf{J}=0$) is assumed. For a Lorentz ($Z=\infty$) plasma the Nernst velocity down a temperature gradient is:
\begin{equation}
	\mathbf{v}_N=\frac{\langle\mathbf{v}\,v^3\rangle}{2\langle v^3\rangle} \label{eq:E_kinetic_origin_nernst}
\end{equation}
	This represents the $v^3$ weighting of the electron velocity $\mathbf{v}$; the field is frozen into the electrons, weighted by how long they keep their direction (time between collisions scales with $v^3$).\cite{hainesHeatFluxEffects1986}

	The numerator of equation \ref{eq:E_kinetic_origin_nernst}  is a moment of $\mathbf{f}_1$, i.e. a flux:
	\begin{equation}
	\langle \mathbf{v}\,v^3\rangle=\frac{4\pi}{3n_e}\int_0^\infty v^6\,\mathbf f_1\,\mathrm dv
	\end{equation}
	The denominator of equation \ref{eq:E_kinetic_origin_nernst} is simply normalization to form the Nernst velocity, and is given through a moment of $f_0$: $\langle v^3\rangle\equiv(4\pi/n_e)\int_0^\infty f_0\,v^{5}\,\mathrm dv$.

In 1D the field generation terms are zero, which simply gives the transverse magnetic field evolution (through $-\partial_x E_y$) to be:
\begin{equation}
	\frac{\partial B_z}{\partial t}= -\frac{\partial v_{N,x} B_z}{\partial x}
\end{equation}

\subsection{Anisotropy and stress from $\underline{\underline{f}}_2$}
\label{sec:ohm_f2}

\begin{figure*}[tb]
	\includegraphics[width=\textwidth]{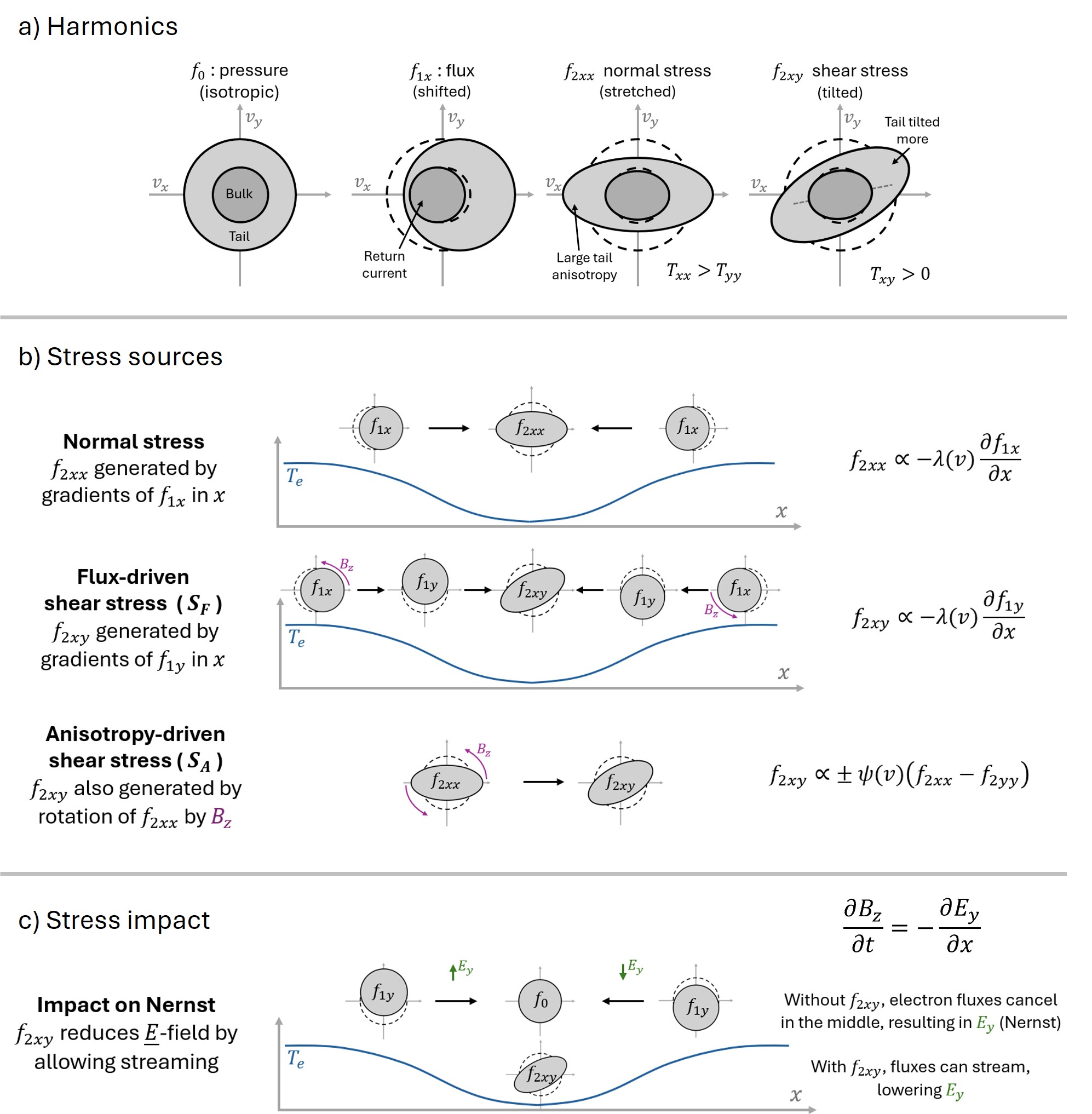}
	\caption{\label{fig:demo} (a) Nature of the electron distribution harmonics, with light grey showing the tail (fast) electrons. (b) The sources of stress at a temperature gradient. (c) The impact of $\underline{\underline{f}}_2$ on Nernst, acting to reduce the electric field ($E_y$) by allowing streaming of electrons where $f_{1y}$ converges.    }
\end{figure*}

Figure \ref{fig:demo}(a) demonstrates the physical nature of the harmonics. In all cartoons the dark grey signifies the bulk of the electron population, while light grey is the high energy tail responsible for carrying flux. The first image is of the isotropic $f_0$. The second is of $f_{1x}$; the electron tail is shifted, signifying a flux to the right. Note that the bulk plasma ($v\approx v_{th}$) forms a return current, as was seen classically in figure \ref{fig:classical_weights}.

$f_{2xx}$ is a stretched distribution. In the case shown, the temperature along the gradient ($T_{xx}$) is larger than the temperature across ($T_{yy}$); this is sausage shaped (prolate). Similarly, $f_{2xx}$ can represent $T_{xx}<T_{yy}$, which is shaped like a pancake (oblate). The first row in figure \ref{fig:demo}(b) shows how this distribution can be formed in a kinetic plasma that is slow to isotropize the electron velocities. An $f_{1x}$ flux of high-energy electrons from a hot region can converge in a cooler region, leading to $T_{xx}>T_{yy}$. More quantitatively, $f_{2xx} \propto -\lambda(v) \partial f_{1x}/\partial x$, where $\lambda(v)$ is the per-speed mean-free path of the electrons. Without magnetization or e-e collisions, $\lambda(v)\propto v^4$.\cite{walsh_paper1}

\begin{figure}[tb]
	\includegraphics[width=\linewidth]{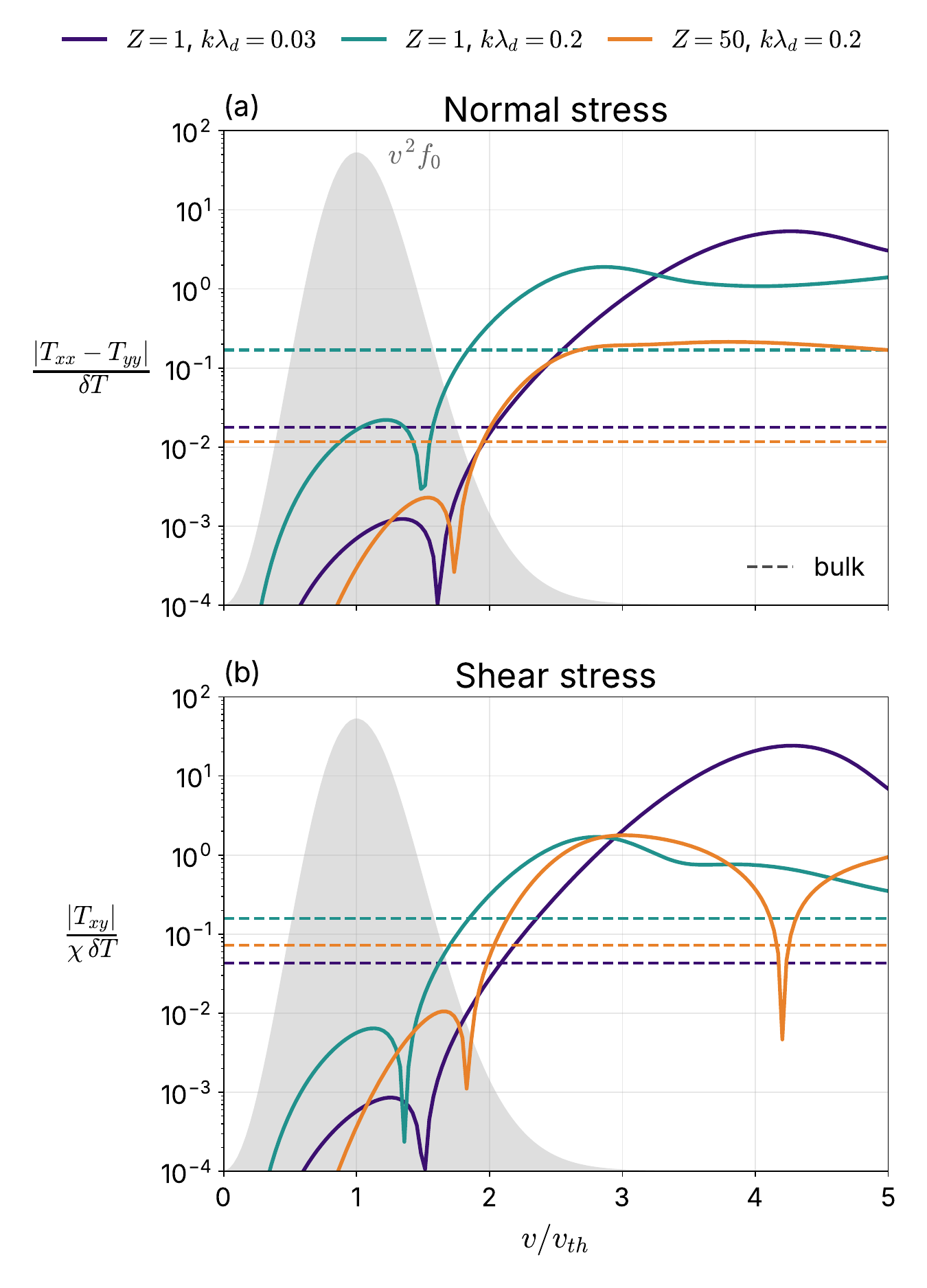}
	\caption{\label{fig:aniso_speed}Temperature anisotropy of a sinusoidal temperature mode, by electron speed (solid) and for
		the plasma as a whole (dashed), per unit $\delta T$. (a) Normal stress $|T_{xx}-T_{yy}|$. (b) Shear stress $|T_{xy}|$ in a
		weak field, per unit $\chi$. The shaded grey region is $v^2 f_0$ for reference, which shows how many electrons are at each speed. The thermal electrons are nearly isotropic; the
		anisotropy sits on electrons above about $2\,v_{th}$, where it exceeds the bulk value by one to two orders of magnitude.
		A shorter wavelength ($k\lambda_d = 0.03 \to 0.2$ at $Z=1$) raises the bulk anisotropy about tenfold and moves its peak
		to slower electrons, as the fastest electrons reach the free-streaming limit. Higher $Z$ ($Z=50$) reduces the normal
		stress about fifteenfold but leaves the shear stress on the $3\,v_{th}$ electrons comparable. These simulations use converged harmonics ($f_6$).}
\end{figure}

Figure \ref{fig:aniso_speed}(a) plots the per-speed anisotropy $|T_{xx}-T_{yy}|/\delta T$; both the plasma ionization ($Z$) and degree of non-locality ($k \lambda_d$) are varied. These simulations are the unmagnetized 1D simulations of the previous work studying heat-flux suppression for a linear $\delta T$ perturbation\cite{walsh_paper1}. The dashed lines show the anisotropy of the bulk electron population, which increases with $k\lambda_d$ and decreases with $Z$; note that this is anisotropy relative to the small temperature perturbation $\delta T$. The important take-away here is that the high energy tail of the distribution has significantly more anisotropy ($>10\times$) than the bulk. In fact, the more non-local cases have less anisotropy in the tail than the more local cases due to tail depletion.

Once a magnetic field is considered, a shear stress ($f_{2xy}$) also emerges in $\underline{\underline{f}}_2$. This is the crucial process for this paper: a tilted cloud of electrons ($T_{xy}\ne0$). Despite the 1D nature of this idealized problem, ($T_{xy}>0$) represents a correlation between electrons moving in $+x$ and $+y$. There are two mechanisms that can form this distribution in 1D.

The first is shown in the second line of figure \ref{fig:demo}(b). It begins with an $f_{1x}$ flux being partially rotated by a magnetic field to make $f_{1y}$; partial rotation means that some of the original $f_{1x}$ flux is retained. In words: there is a flux of electrons in $x$ that have $y$-momentum. Quantitatively, $f_{2xy} \propto \lambda(v)\partial f_{1y}/\partial x$. This process is called flux-driven shear.

The second process generating $f_{2xy}$ is simple: the magnetic field directly rotates an anisotropic $f_{2xx}$ distribution. This is shown in the last row of figure \ref{fig:demo}(b). Quantitatively, $f_{2xy} \propto \psi(v)(f_{2xx}-f_{2yy})$, where $\psi(v)$ is the velocity-dependent Hall parameter. This process is called anisotropy-driven shear.\cite{braginskiiTransportProcessesPlasma1965}

Figure \ref{fig:aniso_speed}(b) plots the normalized shear stress $|T_{xy}|/\chi \delta T$ across an electron population, using the same conditions as figure \ref{fig:aniso_speed}(a). A similar trend is found: the tail electrons have substantially more shear stress than the bulk plasma.

What impact does this shear distribution have on Nernst? It is first worth remembering that Nernst advection originates from the transverse electric field ($E_y$) forcing no net current. With harmonics truncated at $\mathbf{f}_1$, fluxes of electrons towards a temperature minimum (figure \ref{fig:demo}(c)) cancel one another at the center. With $\underline{\underline{f}}_2$ included the electrons are able to stream past one another. For Nernst, the relevant flux is electrons with $y$ momentum traveling in $x$; figure \ref{fig:demo}(c) shows the electric field ($E_y$) required to balance the fluxes at the center where the opposite signed $\mathbf{f}_1$ distributions cancel. $f_{2xy}$ allows these fluxes not to isotropize where they meet, thereby lowering the electric field required to enforce zero current.

Section \ref{sec:ohm_full} quantifies this effect by forming a kinetic Ohm's law that includes shear stress in 1D. The subsequent paper extends this to a generalized shear stress, which also generates magnetic field when more dimensions are considered.

Appendix \ref{sec:aniso_fits} measures the \emph{bulk} normal and shear stress found in these simulations. While the fits are not used in the present work (which is driven by the non-local \emph{tail} stress) the author believes the fits may prove useful for future theoretical work.

\subsection{Full kinetic induction equation}
\label{sec:ohm_full}

With $\underline{\underline{f}}_2$ retained, the $B_z$ field is evolved through:
 \begin{equation}
 	\frac{\partial B_z}{\partial t} = -\frac{\partial (v_{N,x}B_z)}{\partial x} + \frac{2m_e}{5en_J}\frac{\partial \Big\langle v\,\frac{\partial f_{2,xy}}{\partial x}\Big\rangle_w}{\partial x}
 	\label{eq:Bz_kinetic}
 \end{equation}
where $n_J=-\langle \partial f_0/\partial v \rangle_w$ is the number of current-carriers.

The first term in equation \ref{eq:Bz_kinetic} will be referred to as 'pure Nernst', as it is the most common conceptualization of Nernst. The second term is the divergence of the shear stress carried by the current-carrying electrons; its local form, the heat-flux viscosity, was derived by Liu \textit{et al.}\cite{liu2015}. This term has been identified in simulations of reconnection layers\cite{joglekarMagneticReconnectionPlasma2014,liu2017}, where it was found to act in regions of null Nernst velocity. Here it is evaluated generally, across a range of linear wavelengths with no requirement for nulls in Nernst velocity. It is found to be carried by electrons at approximately $3.3v_{th}$, which is higher than any transport process considered in this series.

For a temperature perturbation at wavenumber $k_T$ and a field structure at $k_B$, the response at $k = k_T + k_B$ is
\begin{equation}
	\frac{E_y}{v_N^{c} B} \;=\; \Gamma_{N}(k_T) \;-\; S_F(k_T,k) \;-\; S_A(k_T,k) ,
	\label{eq:closure_form}
\end{equation}
where $v_N^{c}$ is the classical Nernst velocity. $\Gamma_{N}$ is the kinetic suppression function for pure Nernst, which depends on the temperature profile only. $S_A$ is the anisotropy-driven shear stress (generated by the $B$-field rotation of the normal stress) and $S_F$ is the flux-driven shear stress (generated by gradients of $f_{1x}$ in $y$); these sources of stress were highlighted in figure \ref{fig:demo}(b).

To leading order in gradient scale length (small-$k$ limit), with a weak magnetic field:
\begin{equation}
	\Gamma_{N} \simeq 1 - c_N\,(k_T\lambda_d)^2
	\label{eq:smallk1}
\end{equation}
\begin{equation}
	S_F \simeq F\,(k\lambda_d)^2,\qquad
	S_A \simeq A\,(k_T\lambda_d)(k\lambda_d) ,
	\label{eq:smallk2}
\end{equation}
with the coefficients of Table~\ref{tab:nernst_terms}: $c_N = 231$, $F = 90$, $A = 138$ at $Z = 1$ and
$c_N = 301$, $F = 4.9$, $A = 3.9$ at $Z = 100$.

Converting wavenumbers into real-space derivatives for weak non-locality ($k\lambda_d \lessapprox0.05$) and assuming a uniform background, the induction equation becomes:
\begin{align}
	\frac{\partial B_z}{\partial t} = -\frac{\partial}{\partial x}\Big[
	& B_z\Big(v_N^{c} + \lambda_d^2\, c_N\, \frac{\partial^2 v_N^{c}}{\partial x^2}\Big) \nonumber\\
	& + \lambda_d^2\, F\, \frac{\partial^2}{\partial x^2}\big(v_N^{c}B_z\big)
	+ \lambda_d^2\, A\, \frac{\partial}{\partial x}\Big(\frac{\partial v_N^{c}}{\partial x}\,B_z\Big)\Big] .
	\label{eq:realspace}
\end{align}

The corresponding expansion of the heat flow is $q_x/q_x^{c} \simeq 1 - c_q (k_T\lambda_d)^2$, with $c_q = 224$ at $Z=1$ and $248$ at $Z=100$. $c_N/c_q=1.03$ at $Z=1$ and $1.21$ at $Z=100$: pure Nernst follows the heat flow, with the departure from the heat-flow analogy being due to stress.

\section{VFP simulations \label{sec:vfp}}

The simulations use the 1D linear VFP solver outlined in previous work\cite{walsh_paper1}, extended to a structured magnetic
field. The electron distribution is expanded in spherical harmonics up to $f_6$, with the full
e-e collision operator\cite{tzoufrasVlasovFokkerPlanck2011}. Perturbations are Fourier modes about a uniform Maxwellian background of uniform density. All gradients are
along $x$ and the magnetic field is along $z$. The harmonics are assumed quasi-static ($\partial/\partial t = 0$). The electric field is solved for each wavenumber from the zero-current condition. Validation against time-dependent VFP can be found in prior work\cite{walsh_paper1}, as well as in appendix \ref{sec:app_vfp_time} of this paper.  Linearity in $\delta T$ was also confirmed in the previous work\cite{walsh_paper1}.

\subsection{Two modes \label{sec:vfp_twomodes}}
The previous work\cite{walsh_paper1} imposed a temperature mode on a uniform field. Separating Nernst advection from the stress requires the
field to have structure of its own, since the stress terms of Eq.~(\ref{eq:realspace}) depend on the gradient of the field
as well as of the temperature. The problem solved here has two modes,
\begin{equation}
	T = T_0 + \delta T \cos{k_Tx}, \qquad B = B_0 + \delta B\cos{k_Bx}
	\label{eq:two_mode}
\end{equation}
and is linear in $\delta T$ and first order in $\delta B$. As in the previous paper\cite{walsh_paper1}, the temperature perturbation is the slowest decaying eigenmode of the system, not a stationary mode held steady by heating (see appendix \ref{sec:app_vfp_time}). A uniform field ($k_B = 0$) recovers the database from prior work\cite{walsh_paper1}. $E_y$ is read from the field itself, not from a moment of
$\mathbf f_1$.

The current is set
to zero. Amp\`ere's law fixes it at $\mathbf J = \nabla\times\mathbf B/\mu_0$, which is not zero once the field has
structure; neglecting it removes resistive diffusion of the field and the coupling of the resistivity to the
temperature profile (Ohmic heating). For the hot low-density plasmas of interest to this work these are insignificant.

%Relative to the terms retained these are of size
%$1.13\,(\alpha_\perp/\gamma_\perp)(L_T/L_B)(\delta_e/\lambda_{ei})^2$, with $\delta_e = c/\omega_{pe}$ the electron skin depth and
%$\alpha_\perp$, $\gamma_\perp$ the classical resistivity and Nernst coefficients. In a weakly magnetized plasma the
%condition is $(\delta_e/\lambda_{ei})^2 \ll 1$, independent of the field; in a strongly magnetized one it becomes
%$\beta_e \gg 1$, independent of collisionality. Both are satisfied by a wide margin in hot, underdense plasma, where
%$(\delta_e/\lambda_{ei})^2 \sim 10^{-5}$. Because the
%kinetic equation is linear in its drives, a finite current adds a resistive field to the one computed here without
%altering it.

\subsection{Separating terms \label{sec:vfp_separate}}

The terms of equation \ref{eq:realspace} can be separated exactly, due to 1D plasmas in $x$ only having $E_y$ fields by magnetic rotation of the electron distribution. Three solutions are computed for each datapoint.
The first is the full problem. In the second, the field is only allowed to rotate $\mathbf f_1$, with $E_y$ found from the
$\mathbf f_1$ equation alone; the rotated flux never reaches $\underline{\underline{f}}_2$. This is pure Nernst.
In the third solution, the field again rotates only $\mathbf f_1$, but the rotated flux is allowed to drive
$\underline{\underline{f}}_2$: this keeps the flux-driven shear stress ($S_F$) while removing the rotated normal stress
($S_A$ in figure \ref{fig:demo}(b)). Differences between the three give the stress terms of equation \ref{eq:closure_form}.

Appendix \ref{sec:app_vfp_time} discusses and quantifies the difference in measuring a stationary mode (where the temperature perturbation is forced constant) versus a decaying mode (where the temperature is allowed to relax).

\subsection{Database and convergence}

The weak-field set covers 14 values of $Z$ from 1 to 100, 24 values of $k_T\lambda_d$ from 0.002 to 0.4 and 26 values
of $k_B/k_T$ from $-8$ to $12$, a total of 8736 points. The magnetized set covers the same $Z$ and $k$ with 18 values
of $\psi_d$ from 0.1 to 8, a total of $1.5\times10^5$ points. Field structure finer than the temperature ($k_B\lambda_d$ up to 5) is computed for 13 values of $Z$ (585 points, weak field) and for four values of $Z$
and five of $\psi_d$ (360 points). With resolution checks the database contains
$2.2\times10^5$ solutions. Except where otherwise stated, all simulations truncate the harmonic expansion at $f_6$. Example runs at $f_{10}$ confirm convergence at $f_6$; $E_y/(v_N^cB)$ changes by at most 0.010 and leaves pure Nernst unmodified over the range $|k|\lambda_d \le 0.5$.

The velocity-grid error is first order in $1/N_v$: differences between $N_v = 256$, $512$ and $768$ fall in the ratio
3.0 at every datapoint checked. The weak-field set is computed at $N_v = 256$ and $512$ and extrapolated,
$2X_{512} - X_{256}$; the extrapolation changes $E_y/(v_N^cB)$ by at most 0.006, so does not change the message of this work. For the magnetized set the same
correction is computed on a subgrid of eight values of $Z$ and 12 of $k_T$ and interpolated.

\section{Weak Field \label{sec:weak_field}}
% Closure, small-k coefficients and anisotropy closed forms for the Nernst paper (draft, 13 Nov 2026).
% Numbers from closure_v4_constants.json, smallk_v4_table.csv, aniso_closure.py. N_v extrapolated, decaying mode.
% \note{} marks checks; nothing here is final wording.

This section uses VFP datapoints with a weak magnetic field to assess contributions of both pure Nernst and the stress terms on the kinetic induction equation. Section \ref{sec:weak_uniform} begins with a uniform field, explaining the unorthodox result raised in the introduction (figure \ref{fig:nernst_sup}). Pure Nernst is found to follow the heat-flux; it is the stress terms from $f_{2xy}$ that break the heat-flux analogy.

Section \ref{sec:weak_nonuniform} then presents weak-field data when both a temperature and magnetic field perturbation are present. Despite a weak field, $\dot{B}_z$ is highly dependent on the magnetic field profile due to the shear stress terms. For short magnetic field wavelengths relative to the delocalization length, the stress terms are found to reverse the impact of Nernst.

\begin{figure}[tb]
	\includegraphics[width=\linewidth]{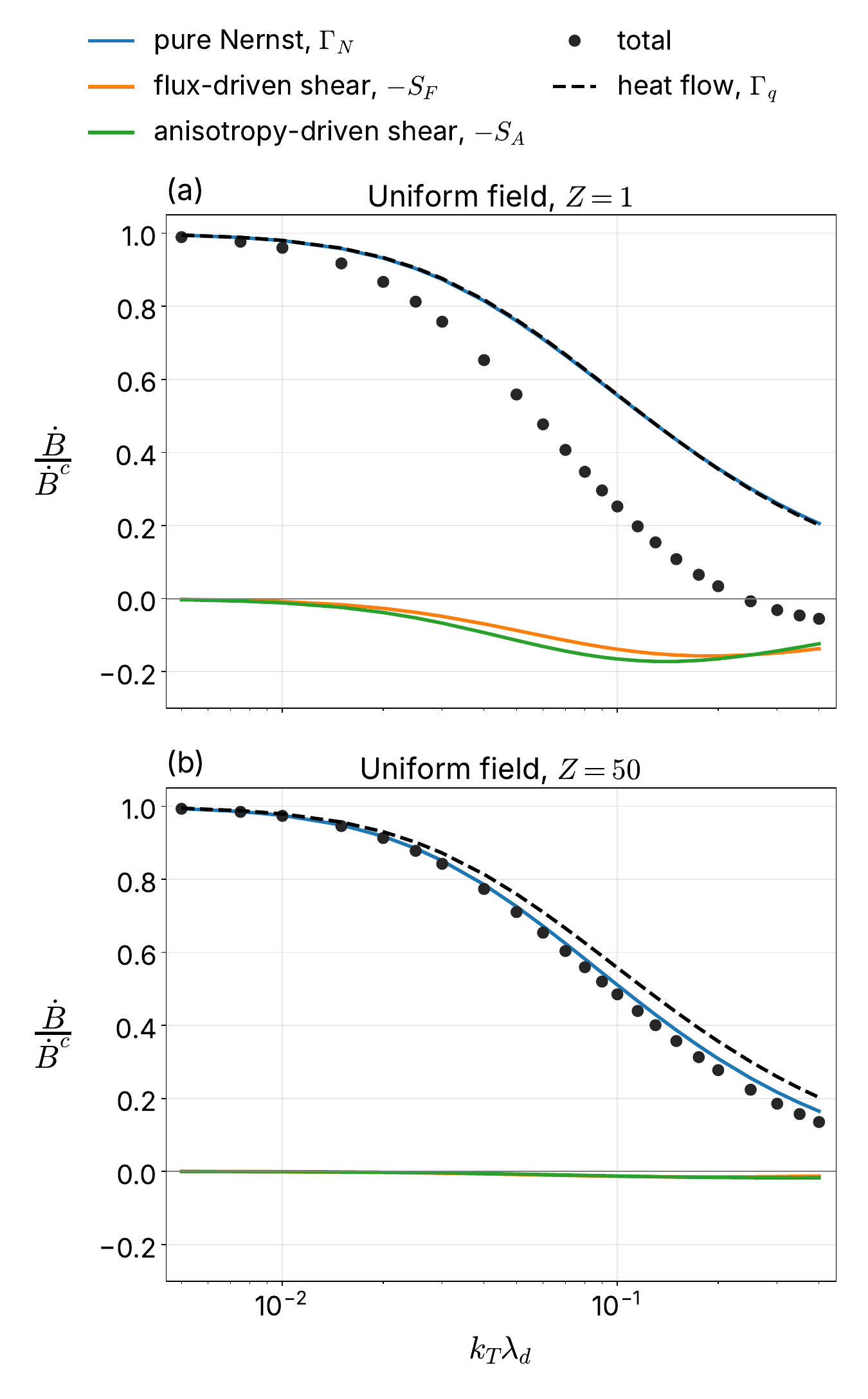}
	\caption{\label{fig:uniform_field_split}  Kinetic suppression of advective B-field transport for a weak uniform field against non-locality ($k\lambda_d$). (a): $Z=1$, (b): $Z=50$. The total suppression (black dots) is split into its different components: pure Nernst advection, flux-driven shear ($-S_F$) and anisotropy-driven shear ($-S_A$). The dashed line is heat-flow suppression $\Gamma_q$, showing that pure Nernst is suppressed by a similar fraction. However, at $Z=1$ the shear terms lower the advection substantially.  }
\end{figure}

\begin{figure*}[tb]
	\includegraphics[width=\textwidth]{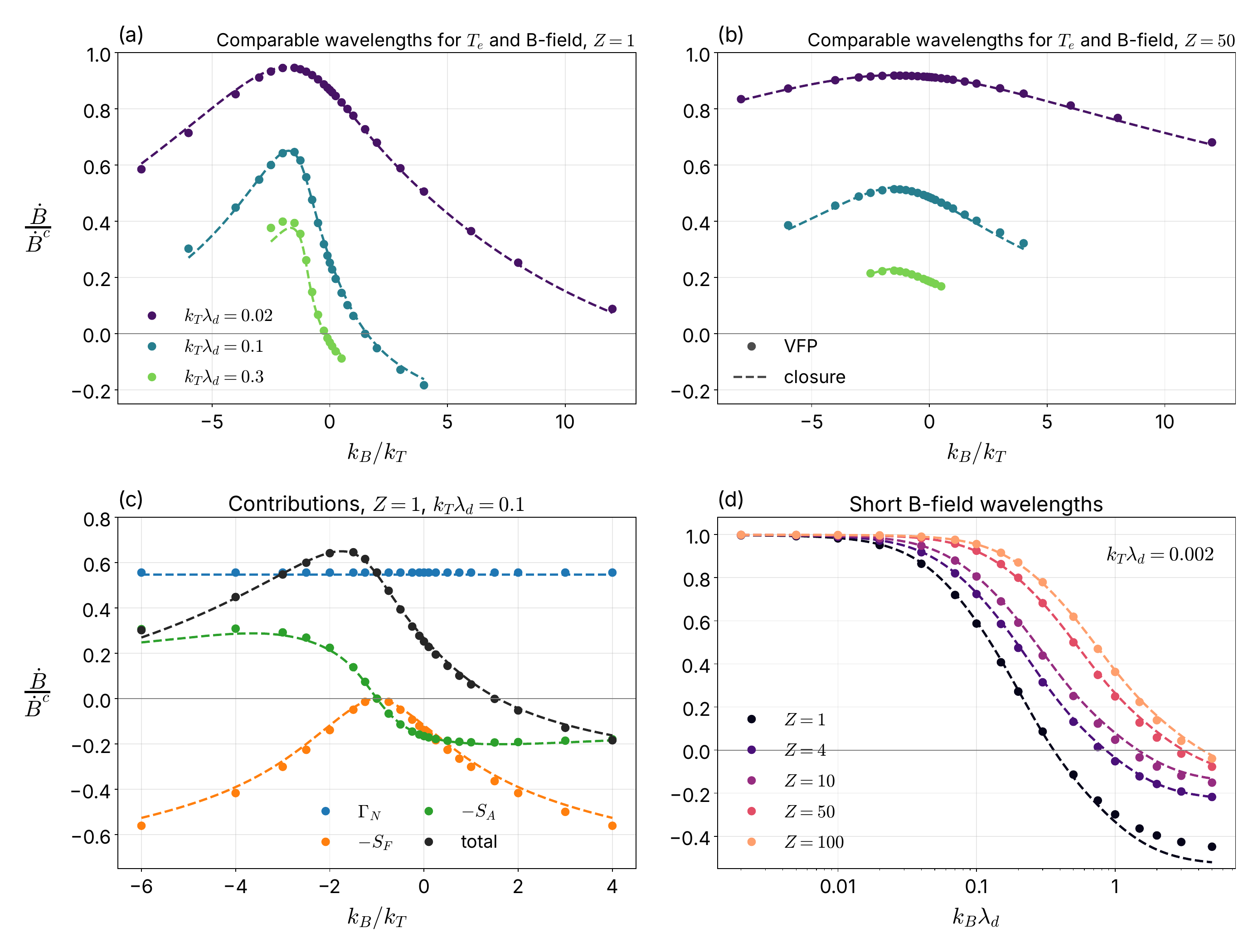}
	\caption{\label{fig:structured_field}  An examination of the B-field structure's impact on kinetic field advection, for weak field strengths. (a,b,c) are for B-field perturbations with comparable wavelengths to the temperature mode. (d) shows how the stress terms can substantially reduce Nernst for short magnetic field wavelengths, despite the field being weak. The plasma is also in a regime usually considered classical ($k_T \lambda_d = 0.002$), with heat-flux being suppressed by less than 1\%.}
\end{figure*}

\subsection{Uniform weak field \label{sec:weak_uniform}}

Figure \ref{fig:uniform_field_split} plots pure Nernst and stress contributions to $\dot{B}_z$ with a uniform field. Figure \ref{fig:uniform_field_split}(a) uses $Z=1$; pure Nernst suppression ($\Gamma_N$) follows the kinetic suppression of heat-flux to within $0.01$. The flux-driven shear and anisotropy-driven shear reduce $\dot{B}_z$ further, with a change of sign occurring above $k_T\lambda_d=0.24$; the magnetic field moves up the temperature gradient.

Figure \ref{fig:uniform_field_split}(b) paints a more familiar picture. Pure Nernst is marginally more suppressed than heat-flux, owing to its marginally higher carrier velocities. The stress terms here are negligible, meaning that the tie between heat-flux and Nernst holds well. The reason for this is not that the magnitude of the stress terms is smaller at $Z=50$; it is smaller \emph{relative} to pure Nernst, with the classical coefficient increasing from $\gamma_\bot^c(Z=1)=0.88$ to $\gamma_\bot^c(Z=\infty)=9.84$.\cite{epperleinPlasmaTransportCoefficients1986,sadlerSymmetricSetTransport2021}

\subsection{Non-uniform weak field \label{sec:weak_nonuniform}}

This section uses the same linear temperature perturbations as section \ref{sec:weak_uniform} but with an added field perturbation at its own wavelength $B_z=\delta B_z \cos(k_B x)$. The field is kept small such that it deflects the electron flux, which produces $E_y$ ($\propto B_z$), but does not yet alter transport along the temperature gradient ($\propto B_z^2$). With a weak uniform offset ($B_z=B_0 + \delta B_z \cos(k_B x)$) the results have been confirmed to be a linear superposition of the uniform field results and the results of this section; this was checked up to $\chi=0.03$, with differences $\le 10^{-3}$. The subsequent section (\ref{sec:mag}) documents how the electron response changes when the background field strength is increased beyond this level.

Traditional wisdom would state that the structure of a weak field should not change the uniform field results. However, figure \ref{fig:structured_field} challenges this wisdom once the stress term is included. The first two panels show the change in magnetic field normalized to classical against $k_B/k_T$. For magnetic fields that are self-generated by Biermann battery, the magnitude of $k_B$ is typically greater than $k_T$\cite{campbell2021measuring,griff-mcmahon2024}, which makes the varied impact of stress highly applicable to magnetized HED experiments.

Figure \ref{fig:structured_field}(a) shows $Z=1$, showing a faster increase in suppression for $k_B/k_T>1$ and for higher levels of non-locality. (b) shows $Z=50$, which has a weaker dependence on the magnetic field perturbation.

Figure \ref{fig:structured_field}(c) takes the $Z=1$, $k_T \lambda_d=0.1$ data and documents the contributions from each term. Pure Nernst, as expected, is independent of magnetic field wavenumber. The anisotropy-driven shear enhances Nernst for $k_B/k_T < -1$ and decreases Nernst for $k_B/k_T > -1$. The flux-driven shear, on the other hand, always decreases Nernst. The explanation for this behavior comes from equation \ref{eq:realspace}; $S_F=F\lambda_d^2 \partial_x^2(v_{Nx} B_z)$ is a diffusion of the Nernst flux over the stress carriers' mean free path, smoothing the advecting flux whatever the phase of the field structure. Meanwhile, anisotropy-driven shear is $S_A=A\lambda_d^2 \partial_x[(\partial_x v_{Nx})B_z]$. The normal anisotropy is set by the convergence of the heat-flux rotated by the field, with the shear's divergence pushing on the field. When the field structure is opposite in phase to the temperature structure, the push lines up with the Nernst flux and enhances it.

Figure \ref{fig:structured_field}(d) takes a plasma that would typically be considered local ($k_T \lambda_d = 0.002$) and plots $\dot{B}/\dot{B}^c$ against $k_B \lambda_{d}$ for a range of $Z$. The result is striking; even though the magnetic field is weak (with Nernst unmagnetized) the change in magnetic field is highly dependent on the B-field wavelength. For long field wavelengths (low $k_B$) every $Z$ returns to the classical result; Nernst is halved at $k_B\lambda_d = 0.12$ for $Z=1$ and at $k_B\lambda_d=0.69$ for $Z=100$. The length-scale that makes stress independent of ionization is $\lambda_{ei}(Z+0.40)/(Z+4.86)$, which is different to the length-scale that makes heat-flux or pure Nernst independent of ionization $\lambda_d = \lambda_{ei}Z/\sqrt{Z+4.45}$.

The results of this section suggest major changes are needed to the existing understanding of Nernst in magnetized HED experiments, particularly for low-$Z$ plasmas (e.g. the gas fill in an ICF hohlraum\cite{abu-shwarebLawsonCriterionIgnition2022}) and for plasmas with finer B-field structure than temperature structure (e.g. from self-generated magnetic fields\cite{campbell2021measuring,griff-mcmahon2024}).

The results also suggest that the concept of a kinetic plasma, typically defined using the temperature length-scale, should be extended to include plasmas with magnetic field structure on the same order as the delocalization length. The conclusion of this paper discusses this in more depth.

While these results are only applicable to the linear regime, section \ref{sec:implication} demonstrates similar effects for a non-linear heat-front with a narrow magnetic field layer.

\section{Magnetization \label{sec:mag}}

In this section an appreciable background $B_z$ magnetic field is applied to the simulations, acting to relocalize transport. This is quantified through the delocalization Hall parameter $\psi_d=13.4 \chi Z/(Z+4.45)$.\cite{walsh_paper1}

\begin{figure}[tb]
	\includegraphics[width=\linewidth]{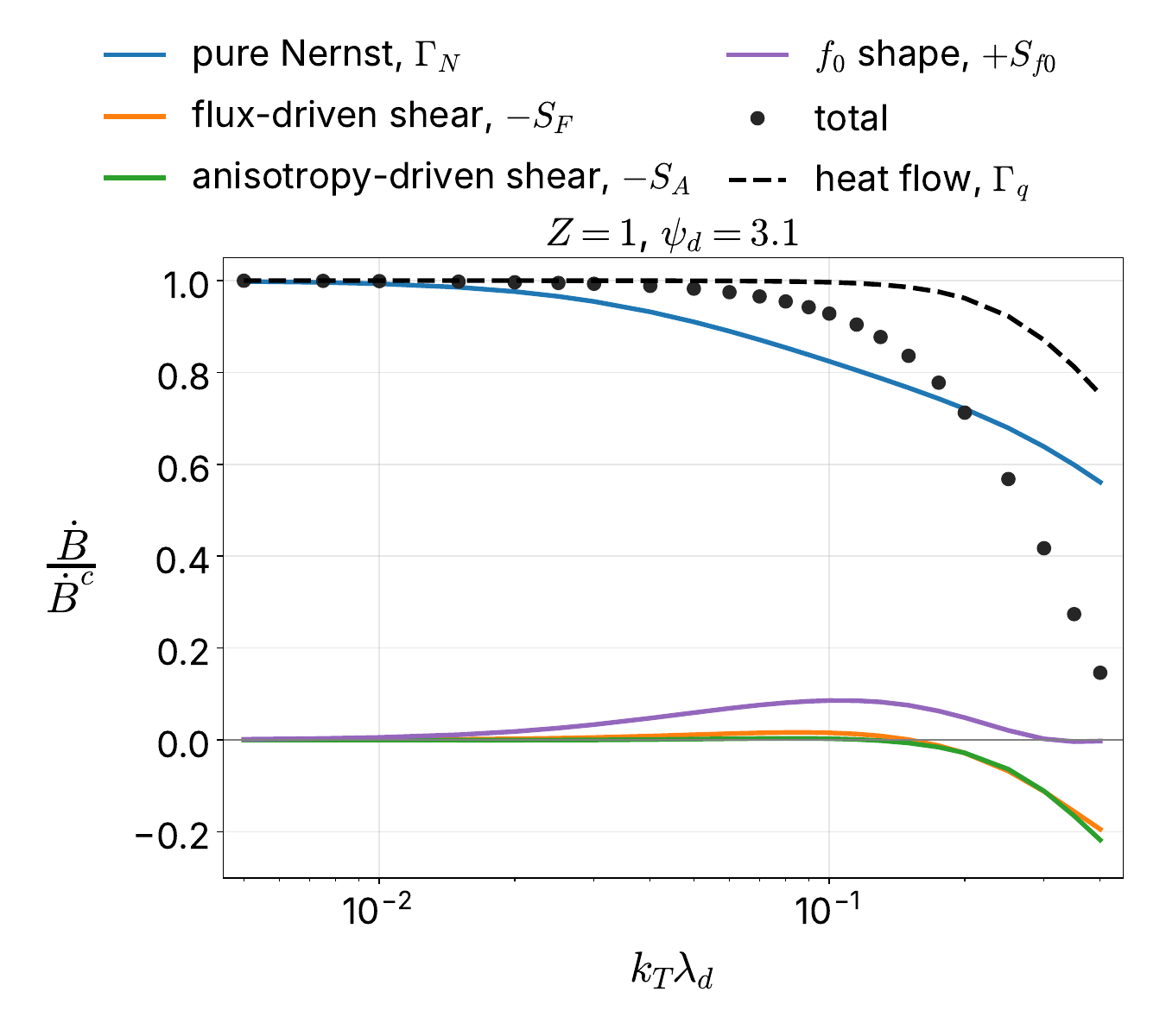}
	\caption{\label{fig:mag_field_split}Same as figure \ref{fig:uniform_field_split}(a), for a magnetized plasma ($Z=1$,
		$\psi_d = 3.1$). At moderate non-locality the stress terms have switched off and the $f_0$-shape term ($S_{f0}$), which exists
		only with a field, raises the total above pure Nernst; pure Nernst itself remains more suppressed than the heat flow. The
		stress returns at large $k_T\lambda_d$.}
\end{figure}

Figure \ref{fig:mag_field_split} uses the same conditions as figure \ref{fig:uniform_field_split}(a), but with a magnetic field corresponding to $\psi_d=3.1$. Pure Nernst relocalizes more slowly than heat-flux; for $k_T \lambda_d=0.1$ the heat-flux is local ($\Gamma_q=0.996$), while pure Nernst is suppressed by 18\%.

Figure \ref{fig:mag_field_split} also shows the total $\dot{B}_z$ rising above pure Nernst for $k_T \lambda_d\lessapprox0.15$. This is not the shear stress, but instead the $f_0$-shape term $S_{f0}$. This is attributed to the magnetic field relocalizing the electron distribution more strongly at the field peaks, affecting the isotropic distortion of $f_0$ directly. The field partly refills the depleted electron distribution tail, but does so non-uniformly across the temperature perturbation. The stress returns at higher non-locality.

\begin{figure}[tb]
	\includegraphics[width=\linewidth]{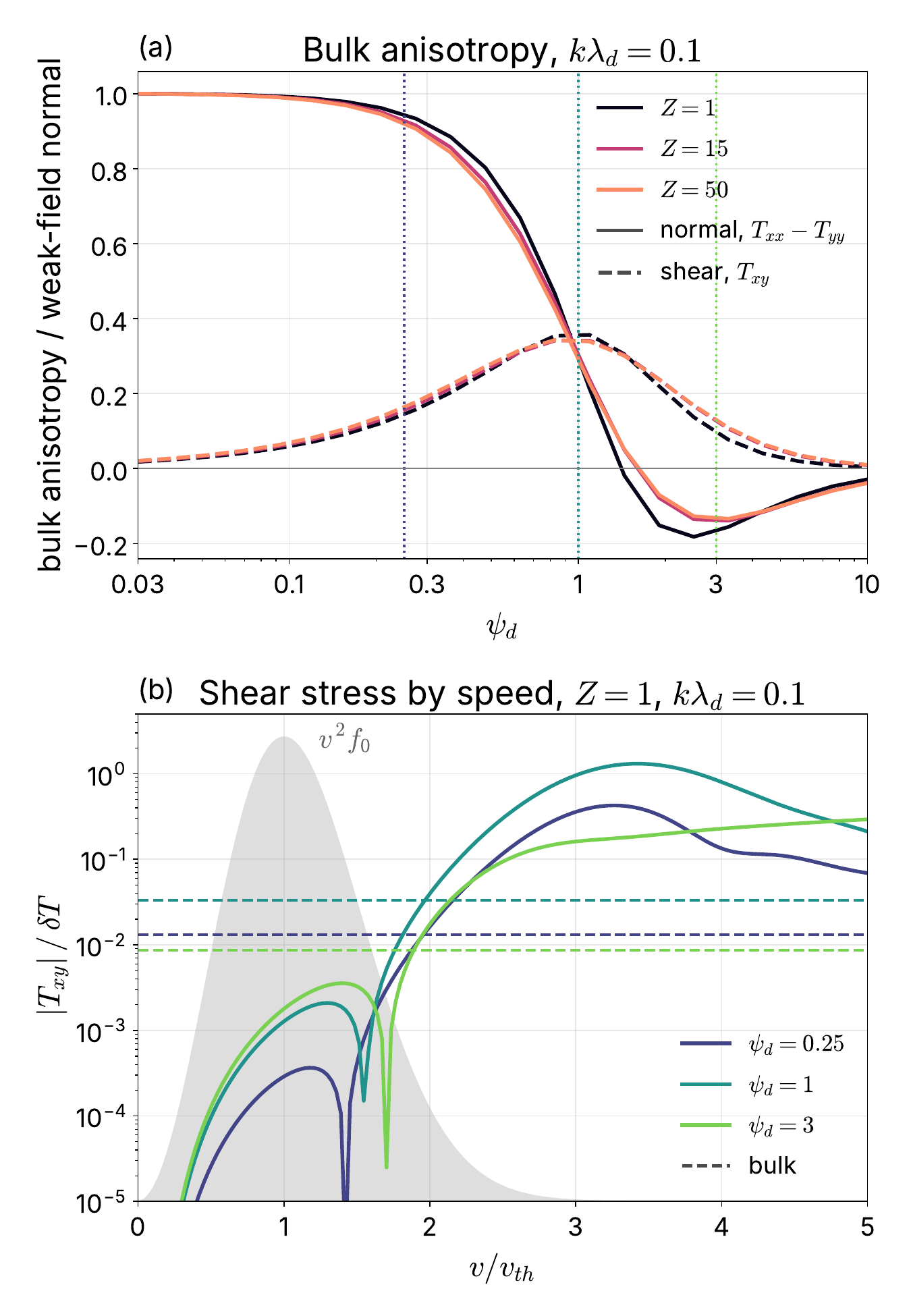}
	\caption{\label{fig:aniso_rotation}The magnetic field rotates the normal stress into a shear stress before suppressing both.
		(a) Bulk normal ($T_{xx}-T_{yy}$, solid) and shear ($T_{xy}$, dashed) anisotropy against $\psi_d$ at $k\lambda_d = 0.1$,
		normalized to the weak-field normal value. All $Z$ collapse onto one curve. The shear stress peaks at about a third of the
		weak-field normal stress near $\psi_d \approx 1$, where the normal stress changes sign as the anisotropy is rotated past
		$45$ degrees. (b) Shear stress by electron speed (solid) and bulk (dashed) at $Z=1$; grey:
		$v^2f_0$. At every field the shear is carried by electrons above about $2\,v_{th}$; past the peak the field moves it to
		faster electrons.}
\end{figure}

Figure \ref{fig:aniso_rotation}(a) plots the bulk normal and shear stress against delocalization Hall parameter as a means to understanding the impact of a magnetic field on the terms originating in $\underline{\underline{f}}_2$. The normal stress ($T_{xx}-T_{yy}$) decreases rapidly for $\psi_d>0.3$, turning negative when the anisotropy has been rotated past 45 degrees (see figure \ref{fig:demo}). At higher $\psi_d$ it recovers back to zero due to relocalization. The shear stress $T_{xy}$ peaks at $\psi_d\approx 1$, as it requires field for rotating the $f_{1y}$ flux, but is also affected by relocalization (note the similarity to Righi-Leduc\cite{walsh_paper1}).

Figure \ref{fig:aniso_rotation}(b) then plots the per-speed shear stress against magnetization for $k\lambda_d=0.1$. The stress carried by the electron distribution tail ($v>2v_{th}$) is initially enhanced by magnetization, before decreasing for $\psi_d>1$.

The stress is relocalized by magnetization at a similar rate to the heat-flux. Table \ref{tab:nernst_terms} gives the Hall parameters of the flux-driven and anisotropy-driven shear components: they are somewhat below $\psi_d$ at $Z=1$ and above it at high $Z$, the latter being consistent with the stress being carried by higher energy electrons ($v\approx3.3v_{th}$) than heat-flux ($v\approx2.6v_{th}$). This contrasts with pure Nernst, which relocalizes more slowly ($0.78\psi_d$).

%\begin{figure}[tb]
%	\includegraphics[width=\linewidth]{mag_correction_floor.pdf}
%	\caption{\label{fig:mag_correction_floor}How a uniform magnetic field removes each kinetic correction ($Z=1$,
%		$k_T\lambda_d = 0.01$). Each correction is normalized to its weak-field value: heat flow $1-q^k/q^c$, pure Nernst
%		$1-E_N^k/E_N^c$ and the stress $S_F+S_A$. The heat-flow and stress corrections are removed together, and both are gone
%		by $\psi_d \approx 2$. Pure Nernst is not fully relocalized: its correction levels off at about a third of its weak-field
%		value. $Z = 15$ and $100$ give the same curves. \note{i think i'll remove this one}}
%\end{figure}

\subsection{Uniform field closure including magnetization}

In a uniform field all terms depend on $k_T$ alone, and the advecting field reduces to one function
(last row of table \ref{tab:nernst_terms}):
\begin{equation}
	\frac{E_y}{v_N^cB} = 1 - S_N\left[1 - P(k_T\lambda_N)\right]
	\label{eq:uniform_closure}
\end{equation}
\begin{equation}
	\lambda_N = 1.15\,\frac{Z+0.67}{Z+0.30}\,\lambda_d, \qquad S_N = 1.03\,\frac{Z+6.32}{Z+4.81},
\end{equation}
which matches the simulations to 0.009 RMS for $k_T\lambda_d \le 0.3$. The stress contribution ($S_U(k_T\lambda_d) \equiv \Gamma_N - E_y/(v_N^cB)$ with $\Gamma_N = P(1.10\,k_T\lambda_d)$) is relocalized
at the same rate as the heat-flow correction, while pure Nernst relocalizes with
$0.78\,\psi_d$. A magnetized uniform field is then approximated by
\begin{equation}
	\frac{E_y}{v_N^cB} \simeq P\!\left(\frac{1.10\,k_T\lambda_d}{1+(0.78\,\psi_d)^{1.6}}\right)
	- S_U\!\left(\frac{k_T\lambda_d}{1+\psi_d^{1.6}}\right),
	\label{eq:uniform_closure_mag}
\end{equation}
which fits the VFP datapoints to 0.03 RMS for $\psi_d \le 3$ (0.06 at $Z=1$); most of the error is the $f_0$-shape term $S_{f0}$, which is omitted here. As in the prior work\cite{walsh_paper1}, $k_T\lambda_d$ must be replaced by a local Knudsen number in a fluid code.

Appendix \ref{sec:app_closure} gives a full closure for the case of a non-uniform magnetic field. Its inclusion in a fluid code is non-trivial, due to the dependence on field structure outlined in section \ref{sec:weak_nonuniform}; this surfaces through the need to calculate Fourier transforms of the temperature and magnetic field profile.

\section{On the relationship between Nernst and heat-flux \label{sec:nernst_heat}}

This section reconsiders the long-held notion of Nernst and heat-flux being two sides of the same coin\cite{hainesHeatFluxEffects1986}. While both are transport down temperature gradients, the author argues that there is just as much similarity between Nernst and Righi-Leduc heat-flow, with both originating from the B-field rotation of $f_{1x}$ flux (Nernst through the $E_y$ field enforcing zero current).

The idealized single carrier model\cite{walsh_paper0} can be used to demonstrate how Righi-Leduc is the rotated heat-flux (accurate to within 12-28\% at weak field):
\begin{equation}
	\kappa_{\wedge}^c \approx \psi_\kappa \kappa_{\bot}^c
\end{equation}
with the extra magnetization factor resulting in Righi-Leduc being carried by a higher energy population than perpendicular heat-flux ($Z=\infty$ peak carrier velocities of $2.90v_{th}$ versus $2.62v_{th}$).

Nernst is read from the same rotated flux, $f_{1y}$, as Righi-Leduc. Remembering that Nernst comes from 'un-rotating' the $\beta_\wedge^c$ coefficient ($\gamma_{\bot}^c=\beta_\wedge^c/\chi$),\cite{walshExtendedmagnetohydrodynamicsDensePlasmas2020,walsh_paper0} the Nernst coefficient can loosely be written:
\begin{equation}
	\gamma_{\bot}^c \sim \frac{\kappa_{\wedge}^c}{\psi_\kappa} \approx \kappa_\perp^c \label{eq:gam_loose}
\end{equation}

A more accurate interpretation comes from the two-carrier model of the classical coefficients\cite{walsh_paper0}, which splits the electron population into a high energy flux driven down a temperature gradient and the low energy return current balancing the flux. The Nernst and heat-flux coefficients then come from a difference in the properties of those two populations.

Nernst is proportional to the difference in energy of the two fluxes ($\Delta u$) times the difference in rotation of the two fluxes ($\Delta h$). Righi-Leduc is also dependent on $\Delta h$, while perpendicular heat-flux is dependent on $\Delta u$.\cite{walsh_paper0}. So, in some senses Nernst is similar to perpendicular heat-flux, and in other senses Nernst is similar to Righi-Leduc.

\begin{figure}[tb]
	\includegraphics[width=\linewidth]{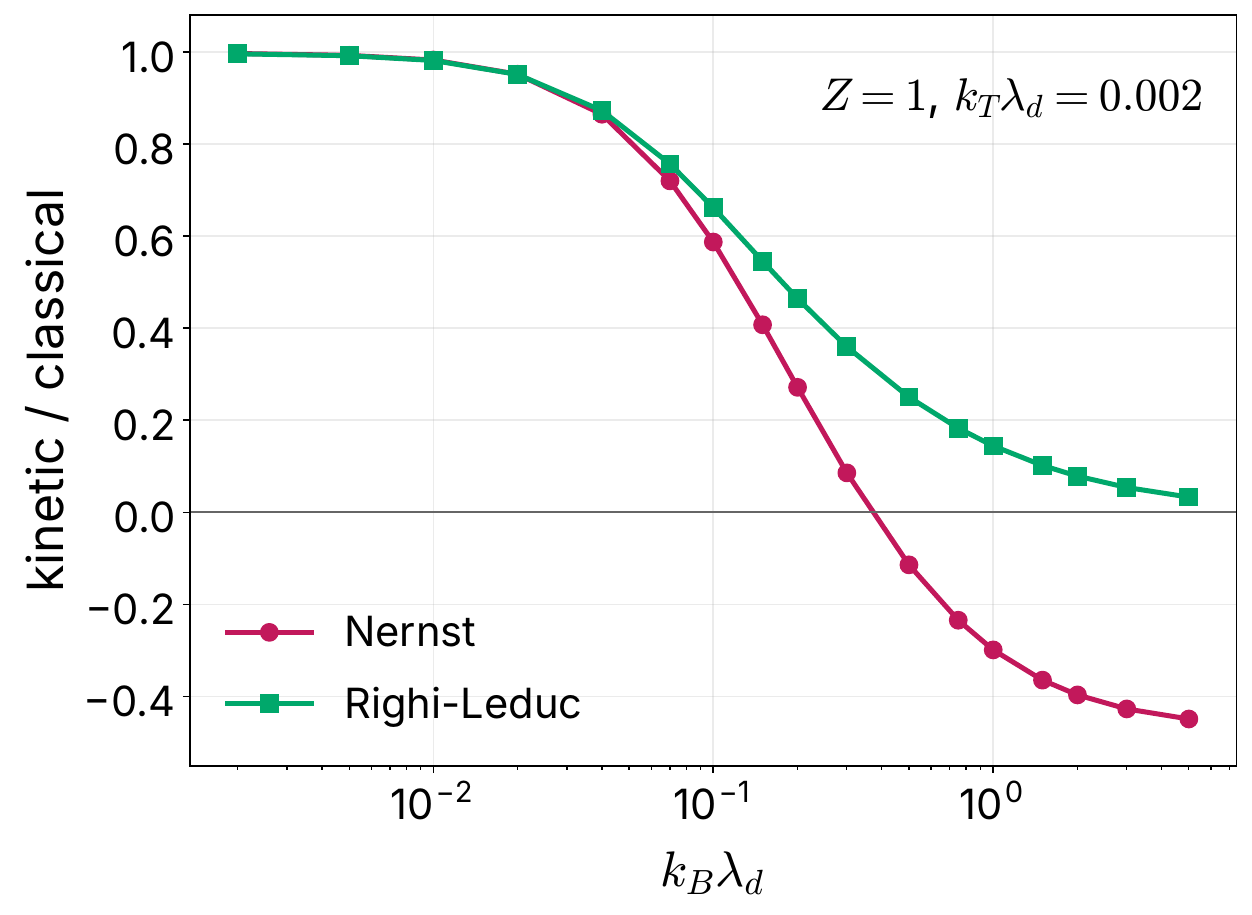}
	\caption{\label{fig:structured_field_d_Z1_RL} Suppression of Nernst and Righi-Leduc at $Z=1$ $k_T\lambda_d=0.002$ for a range of magnetic field wavelengths. Similar to Nernst, Righi-Leduc is suppressed by kinetic stress in non-uniform field profiles, even when it would otherwise be considered classical. }
\end{figure}

As both Righi-Leduc and Nernst are dependent on the rotated flux $f_{1y}$, whose structure is set by both the temperature and magnetic field profiles, Righi-Leduc is \emph{also} reduced by the kinetic stress for short wavelength magnetic field structure, even when the plasma would otherwise be considered local. Figure \ref{fig:structured_field_d_Z1_RL} plots the kinetic suppression of Righi-Leduc and Nernst for a $Z=1$, $k_T \lambda_{d}=0.002$ plasma for a range of weak field perturbations $k_B \lambda_d$. Prior work\cite{walsh_paper1}, which only included temperature perturbations, evaluates Righi-Leduc as being classical across this range; the stress lowers it by over 80\% at $k_B \lambda_{d} = 1$. Righi-Leduc is halved at a similar field structure scale to Nernst ($L_B=2.4\lambda_{ei}$ at $Z=1$ and $15\lambda_{ei}$ at $Z=100$, compared with $3.5\lambda_{ei}$ and $14\lambda_{ei}$ for Nernst), but does not reverse sign. This result suggests that estimates of self-magnetization in ICF hot-spots should be re-evaluated, with previous work finding self-magnetization to be heavily mitigated by Righi-Leduc\cite{walshSelfGeneratedMagneticFields2017,2021,walshNonlinearAblativeRayleighTaylor2023}.

Cross-gradient Nernst ($\gamma_\wedge^c$) completes the picture. Where Nernst originates in $E_y$ canceling the current of the once-rotated flux, cross-gradient Nernst originates in $E_x$ canceling the current of the flux rotated a second time, back along $\nabla T_e$. It is related to Nernst exactly as Righi-Leduc is to perpendicular heat-flux: in the Lorentz limit, both cross terms are their perpendicular partner multiplied by the same carrier Hall parameter\cite{walsh_paper0}, and the cross-gradient-Nernst carriers are correspondingly slightly faster ($2.96v_{th}$) than Righi-Leduc ($2.90v_{th}$).

\section{Implication for a non-linear thermal front\label{sec:implication}}

The results in this paper find Nernst to be heavily suppressed by kinetic effects far beyond the heat-flux analogy. But do the linear results apply to non-linear problems? Will reduced demagnetization by shear stress occur in real experiments?

\begin{figure}[tb]
	\includegraphics[width=\linewidth]{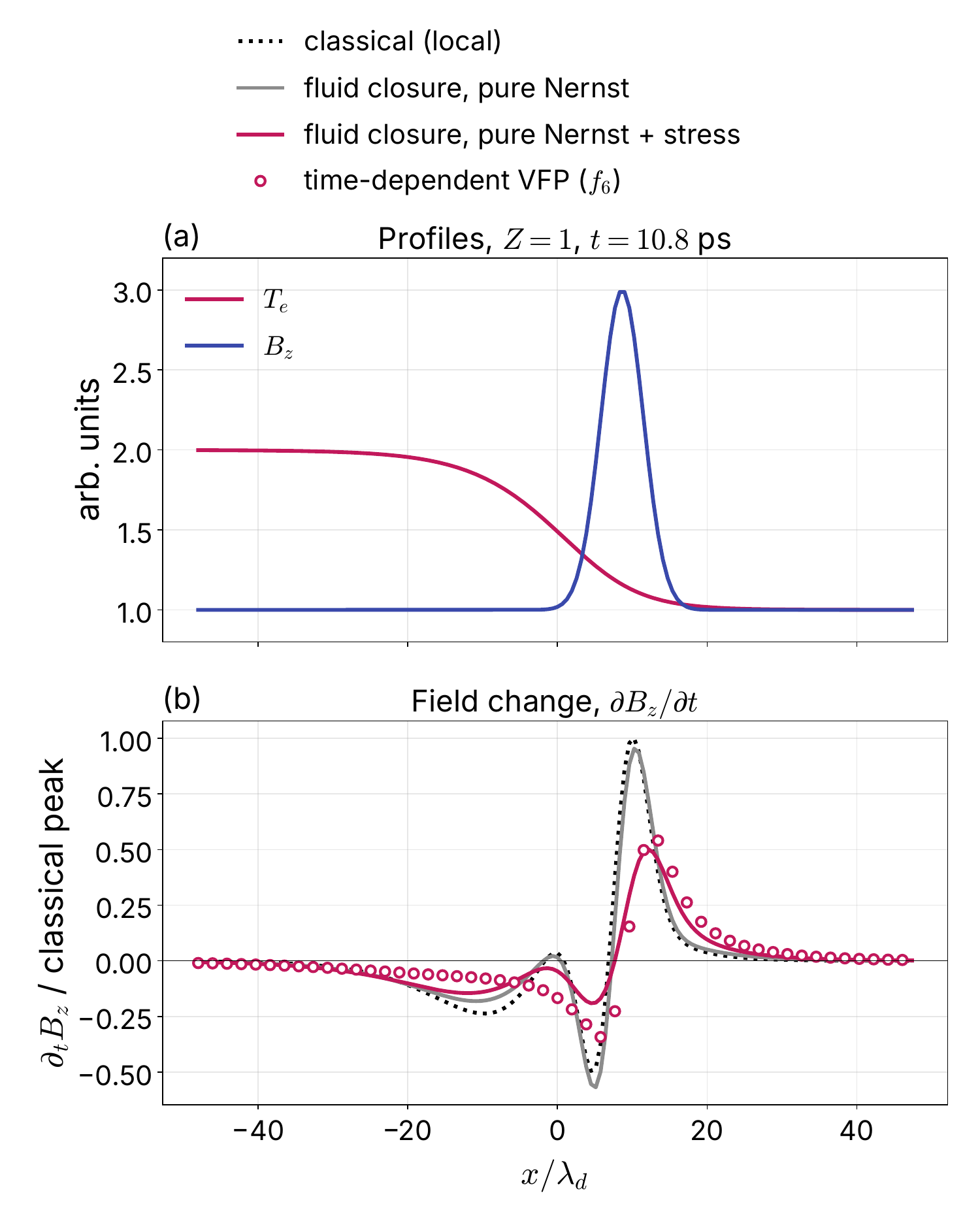}
	\caption{\label{fig:front_merged_Z1} Lowered Nernst demagnetization on a non-linear heat front. (a) temperature and magnetic field profiles. (b) Rate of change of $B_z$ calculated using a full VFP calculation (red circles) and classical (dotted). The grey line is using the pure Nernst fluid closure measured from linear simulations in this paper, showing no reduction from classical. Once the stress term is included (red line) there is surprising agreement with the full VFP simulations.  % resolved 19 Nov: legend now reads 'pure Nernst'
	}
\end{figure}

A time-dependent 1D VFP calculation\cite{walsh_paper1,sherlockSuppressionBiermannBattery2020} is run for a non-linear heat-front. The temperature profile in figure \ref{fig:front_merged_Z1} is evolved until the advection of magnetic field reaches a steady state. The plasma is hydrogen ($Z=1$) with ions kept stationary. The initial temperature profile goes from  $T_e = 1$ to $0.5$\,keV over a $\tanh$ profile of width $16\lambda_d$ ($\lambda_d$
being evaluated at the hot temperature, where $\lambda_{ei} = 21.9\,\mu$m and $\tau_{ei} = 1.17$\,ps). The magnetic field is
imposed and held static: $B_z = B_0\left[1 + 2\exp\!\left(-(x-x_0)^2/(4\lambda_d)^2\right)\right]$, centered at $9.6\lambda_d$
down the gradient from the front, with $B_0$ giving $\chi = 0.03$ at the hot temperature (weak
magnetization). The domain is $2.4$\,mm with 800 cells, and velocity space uses 192 points to $6\,v_{th}$. The field evolution $\partial B_z/\partial t =
-\partial E_y/\partial x$ is evaluated at $t = 10.8$\,ps, once $\mathbf f_1$ has relaxed. The fluid closure (presented in appendix \ref{sec:app_closure}) is evaluated on the VFP temperature profile at the
same time, with its filter lengths set by $\lambda_d$ at the temperature of the field peak.

The solid grey line in figure \ref{fig:front_merged_Z1}(b) shows the expectation from the fluid closure using pure Nernst with no stress terms. The magnetic field evolution is essentially classical (dotted line). The full fluid closure built from linear theory in this paper is plotted as a solid red line, with $\approx50\%$ less change in field. The fully non-linear time-dependent VFP simulations (red circles) are remarkably close to the linear fluid closure when the stress term is included. This validates the approach of this paper and suggests kinetic stress does play a role in enhancing magnetization across magnetized-HED experiments.

\section{Conclusion}

This work studies magnetic field advection due to Nernst in the linear regime using $>2.2\times10^5$ VFP datapoints across ionization, magnetization, temperature perturbation wavelength and magnetic field perturbation wavelength. The electron distribution's $\underline{\underline{f}}_2$ harmonic is found to be crucial in allowing electron streams to pass through one another without losing their $y$-momentum ($f_{2xy}$), thereby lowering the effect of Nernst below the heat-flux analogy. This process is the kinetic version of the heat-flux viscosity\cite{liu2015}, referred to here as the kinetic stress.

The magnetic field profile is found to play an important role in magnetic transport, even when the field is too weak to suppress the classical coefficient ($\psi_\gamma \ll 1$\cite{walsh_paper0}) or relocalize transport ($\psi_d \ll 1$\cite{walsh_paper1}). This is due to the Nernst electric field depending on the deflection direction of electrons in a magnetic field ripple. Only for out of phase field perturbations ($k_B=-k_T$) does B-field advection not depend on the kinetic stress.

For electron populations traveling further between collisions than the perturbation wavelength, Nernst is suppressed. This is true even when the plasma would otherwise be considered to undergo classical transport ($k_T \lambda_d =0.002$). This can be understood through considering that an electron will be deflected back and forth by the magnetic field before it collides; a local approximation (or a kinetic simulation truncating at $\mathbf{f}_1$) assumes the electron is deflected in one direction along its whole path.

Recent work\cite{walsh_paper1} defined a revised Knudsen number for assessing kinetic heat-flow suppression ($Kn_d = 8.28 \lambda_{d\bot}/L_T$). This work suggests a field-structure Knudsen number should also be employed:
\begin{equation}
	Kn_B = 14.8 \frac{Z+0.40}{Z+4.86}\frac{\lambda_{ei}}{L_B} \label{eq:KnB}
\end{equation}
where the prefactor has been chosen such that $Kn_B=1$ corresponds to a 50\% reduction in Nernst due to kinetic stress. $L_B$ is the magnetic field structure scale ($L_B = 1/k_B$ for a sinusoidal perturbation), which is poorly approximated by $|\mathbf{B}|/|\nabla|\mathbf{B}||$ when there is a field ripple on top of a background field. Equation \ref{eq:KnB} assumes a weak magnetic field and $Kn_d \ll 1$. Righi-Leduc heat-flow is halved at a similar $Kn_B$.

While this work is formulated in the linear regime, the results are shown to apply to a weakly magnetized non-linear heat-front. While a kinetic treatment of 'pure Nernst' anticipates little change to the field profile from kinetics, the kinetic stress reduces field advection by $\approx 50\%$.

A more complete demonstration of kinetic stress in magnetized-HED experiments will be the subject of future work. By lowering Nernst more than heat-flux, demagnetization is expected to occur more slowly than current predictions, particularly for low-$Z$ plasmas. The results in this paper could explain why arbitrary multipliers on the Nernst term (e.g. $0.1\times$\cite{leal2025}) are often used to match measurements of electron temperature in ICF hohlraums.

This is the third paper in a series studying plasma transport through the lens of the electron populations that carry the flux of magnetic field and heat\cite{walsh_paper0,walsh_paper1}. The final paper will apply this approach to magnetic field generation mechanisms, including a generalized form for the kinetic stress in Ohm's law.

% resolved 19 Nov (sentence added in Sec. Two modes): not stationary modes. Change text

% resolved 19 Nov (13.4 in both places): psi d constant is wrong

% resolved 19 Nov (two definitions, now stated): carrier speed numbers are wrong

\begin{acknowledgments}
	This work was performed under the auspices of the U.S. Department of Energy by Lawrence Livermore National Laboratory under Contract DE-AC52-07NA27344.

	This document was prepared as an account of work sponsored by an agency of the United States government. Neither the United States government nor Lawrence Livermore National Security, LLC, nor any of their employees makes any warranty, expressed or implied, or assumes any legal liability or responsibility for the accuracy, completeness, or usefulness of any information, apparatus, product, or process disclosed, or represents that its use would not infringe privately owned rights. Reference herein to any specific commercial product, process, or service by trade name, trademark, manufacturer, or otherwise does not necessarily constitute or imply its endorsement, recommendation, or favoring by the United States government or Lawrence Livermore National Security, LLC. The views and opinions of authors expressed herein do not necessarily state or reflect those of the United States government or Lawrence Livermore National Security, LLC, and shall not be used for advertising or product endorsement purposes.

	The data that support the findings of this study are available from the corresponding author upon reasonable request.
\end{acknowledgments}

\appendix
\section{Bulk anisotropy and shear fits \label{sec:aniso_fits}}

% Normal and shear anisotropy closed forms (N_v extrapolated, decaying mode).
% Set for a single column (two-column PRE/PoP width, 3.375 in); checked free of overfull boxes.

Both fits take the same form: the small-$k$ law, which grows as $(k\lambda_{ei})^2$, multiplied by a sum of three
filters that capture its roll-off at shorter wavelengths. The bulk normal anisotropy, per unit $\delta T$, is
\begin{equation}
	\frac{T_{xx}-T_{yy}}{\delta T} = -a(Z)\,(k\lambda_{ei})^2
	\sum_{j=1}^{3}\frac{w_j}{1+(s_jk\lambda_d)^2},
	\label{eq:aniso}
\end{equation}
with
\begin{equation}
	a(Z) = \frac{10395}{144}\,\frac{Z^2+0.0363Z}{Z^2+7.432Z+8.211}.
	\label{eq:aniso_a}
\end{equation}
The weights $w_j$ and lengths $s_j$ depend on $Z$ through $\zeta = 1/(1+Z/3.979)$:
\begin{align}
	w_1 &= 0.0980+0.0235\,\zeta, & s_1 &= e^{0.7127+0.7240\,\zeta}, \nonumber\\
	w_2 &= 0.4134+0.1811\,\zeta, & s_2 &= e^{2.0848+0.4536\,\zeta}, \label{eq:aniso_ws}\\
	w_3 &= 1-w_1-w_2,            & s_3 &= e^{3.2230+0.1985\,\zeta}. \nonumber
\end{align}
Equation~(\ref{eq:aniso}) is within 1.3\% of the simulations for $Z = 1$ to 100 and
$k\lambda_d \le 0.5$ (0.33\% RMS). In the Lorentz limit $a \to 10395/144 = 72.19$, the value given by
$\underline{\underline{f}}_2 = -(2/9)(v/\nu)^2\nabla\nabla f_0$ at zero current\cite{burnett1936,liu2015}.

A weak magnetic field rotates the normal anisotropy into a shear anisotropy, which is first order in $\chi$:
\begin{equation}
	\frac{T_{xy}}{\chi\,\delta T} = -a_S(Z)\,(k\lambda_{ei})^2
	\sum_{j=1}^{3}\frac{w^S_j}{1+(s^S_jk\lambda_d)^2},
	\label{eq:shear}
\end{equation}
with
\begin{equation}
	a_S(Z) = 1123\,\frac{Z^2-0.343Z}{Z^2+11.99Z+43.68},
	\label{eq:shear_a}
\end{equation}
and, with $\zeta_S = 1/(1+Z/6.254)$,
\begin{align}
	w^S_1 &= 0.0019-0.0084\,\zeta_S, & s^S_1 &= e^{-0.3078+0.9966\,\zeta_S}, \nonumber\\
	w^S_2 &= 0.2888+0.1543\,\zeta_S, & s^S_2 &= e^{2.1222+0.6132\,\zeta_S}, \label{eq:shear_ws}\\
	w^S_3 &= 1-w^S_1-w^S_2,          & s^S_3 &= e^{3.3284+0.1432\,\zeta_S}. \nonumber
\end{align}
Over the same range, Eq.~(\ref{eq:shear}) is within 1.9\% of the simulations for $k\lambda_d \le 0.15$ and
3.3\% above (1\% RMS).

The shear is more sensitive to ionization than the normal anisotropy: $a_S$ rises 87-fold from $Z = 1$ to the
Lorentz limit, against 16-fold for $a$. It also rolls off faster with $k$: at $k\lambda_d = 0.1$ it is 0.19 to
0.25 of its small-$k$ law, against 0.37 to 0.41 for the normal anisotropy.

\begin{figure*}[tb]
	\includegraphics[width=\textwidth]{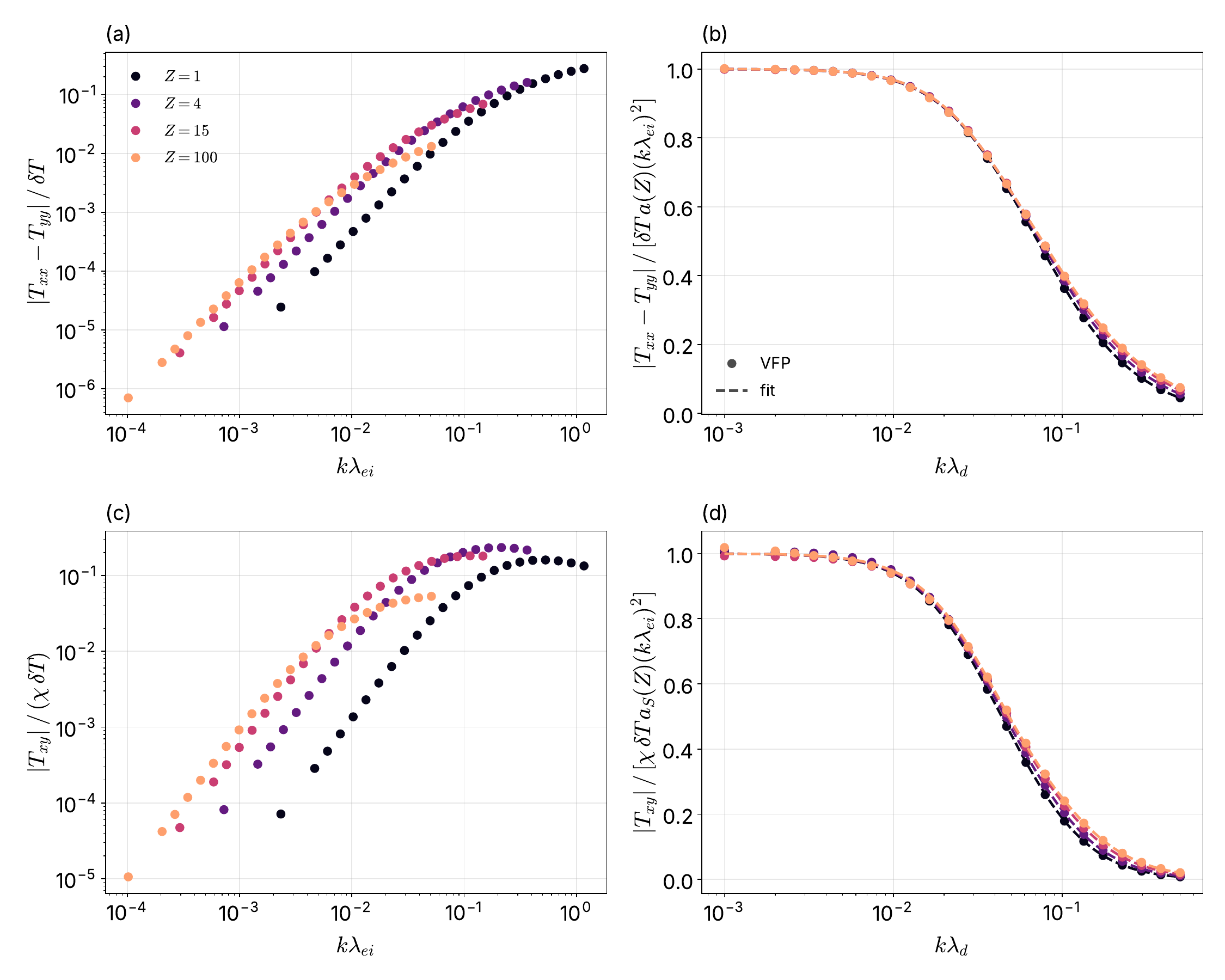}
	\caption{\label{fig:aniso} Normal (a, b) and shear (c, d) stress anisotropy driven by a
		temperature perturbation of wavenumber $k$. (a, c) VFP simulations against $k\lambda_{ei}$.
		(b, d) The same, normalized by the small-$k$ law, against $k\lambda_d$; dashed lines are
		Eqs.~(\ref{eq:aniso}) and (\ref{eq:shear}).}
\end{figure*}

\section{Comparison with time-dependent VFP and mode definition \label{sec:app_vfp_time}}

\begin{figure}[tb]
	\includegraphics[width=\linewidth]{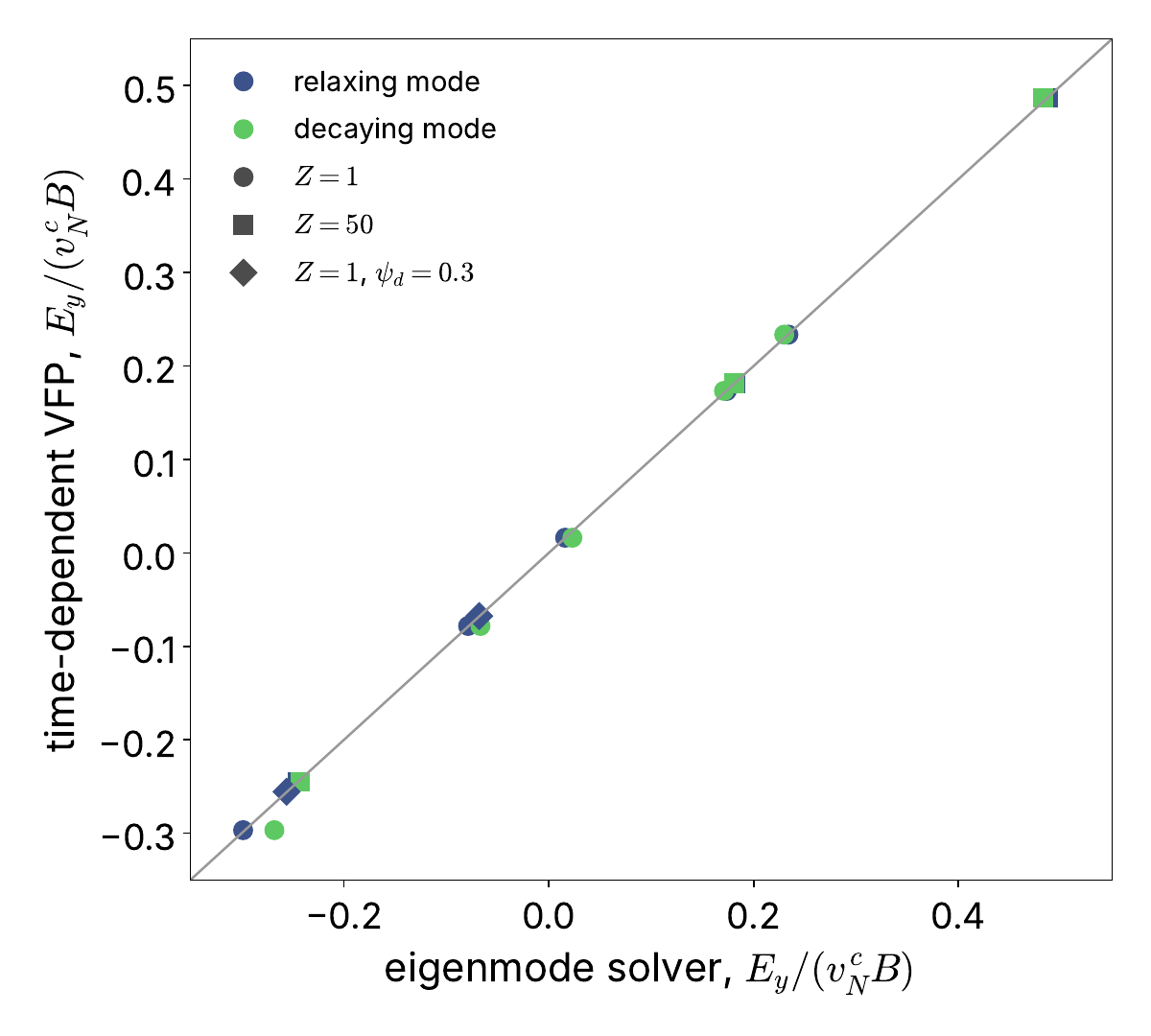}
	\caption{\label{fig:figA_time_dependent}Time-dependent VFP (symbols) against the linear eigenmode solver (lines) for the
		two-mode problem.}
\end{figure}

The linear solver and its validation against time-dependent VFP are described in prior work\cite{walsh_paper1}. As there, the
temperature perturbation is the slowest decaying eigenmode, with the higher harmonics quasi-static. A stationary
definition, in which the mode is held steady by heating, differs from the decaying one by 0.012 RMS (maximum 0.06) in
$E_y/(v_N^cB)$ for a weak field. For the two-mode problem the comparison is repeated in Fig.~\ref{fig:figA_time_dependent}:
a time-dependent VFP calculation with a static field ripple and a freely decaying temperature mode reproduces the
eigenmode solution to within 0.6\% of the classical value at $Z = 1$ and 50, including the sign reversal for field
structure finer than the temperature.

\section{Closure for structured fields \label{sec:app_closure}}

The non-linear front of Sec.~\ref{sec:implication} uses the following closure for a structured field. Its derivation and
generalization are left to future work. With $k = k_T + k_B$ and $H(x) = 1/(1+x^2)$,
\begin{align}
	\frac{E_y}{v_N^cB} &= \Gamma_N - S_F - S_A, \nonumber\\
	\Gamma_N &= \sum_{j} w^N_j\, H\!\left(\ell_j k_T\lambda_d\,\tfrac{Z+4.986}{Z+5.729}\right), \nonumber\\
	S_F &= F\,(k\lambda_{ei})^2 \sum_{j} w^F_j\, H(x_j)\, H(y_j/r_F), \nonumber\\
	S_A &= A\,(k_T\lambda_{ei})(k\lambda_{ei}) \sum_{j} w^A_j\, h_j\,(1 - g\,x_j y_j h_j/r_A), \nonumber
\end{align}
with $x_j = \ell_j k_T\lambda_d$, $y_j = \ell_j k\lambda_{ei}Z/(Z+5.522)$ and $h_j = H(x_j)H(y_j/r_A)$. Each $H$ is a
Helmholtz filter, so in real space every term is evaluated by smoothing over the lengths $\ell_j\lambda_d$. The constants
are in Table~\ref{tab:closure}. Over $|k|\lambda_d \le 0.5$ the closure matches the simulations to 0.006 RMS. The filter form of $\Gamma_N$ used here is within 0.02 of the simpler $P(1.10\,k_T\lambda_d)$ of Table~\ref{tab:nernst_terms} for $k_T\lambda_d\le0.3$.

\begin{table}[h]
	\caption{\label{tab:closure} Constants of the structured-field closure.  $r_F = 0.6509$, $r_A = 0.8388$, $g = 0.887$;
		$F = 518.4\,\frac{Z^2+0.139Z}{Z^2+11.93Z+22.76}$ and $A = 397.6\,\frac{Z^2+0.139Z}{Z^2+7.183Z+9.665}$.}
	\begin{ruledtabular}
		\begin{tabular}{lccc}
			& $j=1$ & $j=2$ & $j=3$ \\
			\hline
			$\ell_j$ & 27.05 & 8.896 & 2.114 \\
			$w^N_j$  & 0.3073 & 0.4686 & 0.2241 \\
			$w^F_j$  & 0.7749 & 0.2193 & 0.005814 \\
			$w^A_j$  & 0.8120 & 0.1802 & 0.0079 \\
		\end{tabular}
	\end{ruledtabular}
\end{table}

	\bibliographystyle{aipnum4-2}
	\bibliography{library}

\end{document}